\documentclass[twocolumn,reprint,aps,superscriptaddress,longbibliography]{revtex4-2}
\usepackage{amsmath}
\usepackage{amssymb}
\usepackage{amsfonts}
\usepackage[dvips]{graphicx}
\usepackage{subfigure}
\usepackage{dcolumn}
\usepackage{txfonts}
\usepackage{bm}
\usepackage{makeidx}
\usepackage{color}
\usepackage{mathtools}
\usepackage{threeparttable}
\usepackage[colorlinks,linkcolor=blue,anchorcolor=blue,citecolor=blue,urlcolor=blue]{hyperref}
\usepackage{lipsum}
\usepackage{braket}

\begin{document}


\title{Control of quantum coherence of photons exploiting quantum entanglement}

\author{Dianzhen Cui}
\affiliation{Center for Quantum Sciences and School of Physics, Northeast Normal University, Changchun 130024, China}

\author{Xi-Lin Wang}
\affiliation{National Laboratory of Solid State Microstructures, School of Physics, Nanjing University, Nanjing 210093, China}

\author{X. X. Yi}
\affiliation{Center for Quantum Sciences and School of Physics, Northeast Normal University, Changchun 130024, China}

\author{Li-Ping Yang}
\email{lipingyang87@gmail.com}

\affiliation{Center for Quantum Sciences and School of Physics, Northeast Normal University, Changchun 130024, China}

\begin{abstract}
Accurately controlling the quantum coherence of photons is pivotal for their applications in quantum sensing and quantum imaging. Here, we propose the utilization of quantum entanglement and local phase manipulation techniques to control the higher-order quantum coherence of photons. By engineering the spatially varying phases in the transverse plane, we can precisely manipulate the spatial structure of the second-order coherence function of entangled photon pairs without changing the photon intensity distribution of each photon. Our approach can readily be extended to higher-order quantum coherence control. These results could potentially stimulate new experimental research and applications of optical quantum coherence.
\end{abstract}

\maketitle

\section{Introduction}
Three are three main quantum resources of photons having been utilized for quantum-enhanced technologies~\cite{giovannetti2004quantum,dowling2008quantum,lloyd2008enhanced}: quantum states of photons with reduced quantum fluctuations, such as squeezed coherent states~\cite{walls1983squeezed,wu1986generation,slusher1985observation}; quantum entanglement, which characterizes the global quantum correlations between photons, such as N00N states~\cite{kok2002creation,afek2010high} and polarization entanglement~\cite{Kwiat1999ultrabright,Gisin2002quantum}; and higher-order quantum coherence arising from the statistical correlation of electromagnetic fields at multiple space-time points~\cite{Glauber1963}. Spatial correlation and quantum entanglement of photons have been widely used to enhance the signal-to-noise ratio in quantum imaging~\cite{brida2010experimental,ono2013entanglement,gregory2020imaging,moreau2019imaging} and increase the sensitivity in phase measurements with weak quantum light~\cite{wolfgramm2013entanglement,Israel2014supersensitive,he2023quantum}.
Spatial-polarization hyper-entangled photon pairs have been used for quantum holography of complex objects~\cite{Defienne2021polarization} and quantum-enhanced phase imaging~\cite{Camphausen2021quantum,Black23quantum}. Remarkable progress has been made in leveraging both two-body quantum entanglement and the spatial quantum coherence of photon pairs for a variety of valuable applications~\cite{chrapkiewicz2016hologram,Ndagano2022microscopy,zia2023interferometric}. Here, we present a theoretical framework aimed at actively manipulating the spatial structure of higher-order quantum coherence of photons through the utilization of quantum entanglement.

The spatial coherence of a photon pair generated through the spontaneous parametric down-conversion (SPDC) process is primarily determined by the characteristics of the pump laser and the nonlinear media~\cite{walborn2010spatial,Law2004analysis}. The recent advancements in precisely manipulating the transverse spatial properties of photons~\cite{Yu2011Light,devlin2017Arbitrary,Shen2019,forbes2020structured} have provided us with powerful tools to further control the quantum coherence of photon pairs after they exit the source. Transfer of entanglement between the spin and orbital angular momentum (OAM) degrees of photons has been achieved using techniques such as a q-plate, a spatial light modulator (SLM), or a structured metasurface~\cite{Nagali2009quantum,stav2018quantum}. However, previous research has mainly focused on the global entanglement between the two photons. The precise manipulation of the detailed spatial structure of second-order quantum coherence in two-photon states is particularly intriguing, as it has the potential to unlock valuable quantum resources within structured photon pairs for quantum imaging~\cite{brida2010experimental,morris2015imaging,lemos2014quantum,magana2019quantum} and quantum sensing~\cite{Lavery2013detection,Korech2013,Chen2019Quantum,qiu2022fragmental}.

In previous work~\cite{yang2022quantum}, the coherence function of entangled a vortex pair has shown to be modulated by the helical phases of the photons. Recently, this effect has been experimentally demonstrated~\cite{zia2023interferometric,gao2023full,huang2023manipulating}. In this paper, we propose a general theoretical frame for the precise manipulation of higher-order quantum coherence of photons. By engineering the phases of the photons in their transverse plane, we can tailor the spatial structure of the quantum coherence function of photon pairs on-demand, while keeping the photon number density distribution unchanged. This method can be applied for higher-order quantum optical coherence control. Our results could inspire captivating experiments and new applications that harness the quantum coherence of photons.

The article is organized as follows. In Sec.~\ref{sec2}, we present the method to control the spatial structure of the second-order quantum coherence of entangled photon pairs. In Sec.~\ref{sec3}, we demonstrate the coherence control of product-state photon pairs via Hong-Ou-Mandel (HOM) interference. Manipulation of higher-order coherence of photons is elucidated in Sec.~\ref{sec4}. The conclusions are summarized in Section~\ref{sec5}.

\begin{figure*}
\includegraphics[width=14cm]{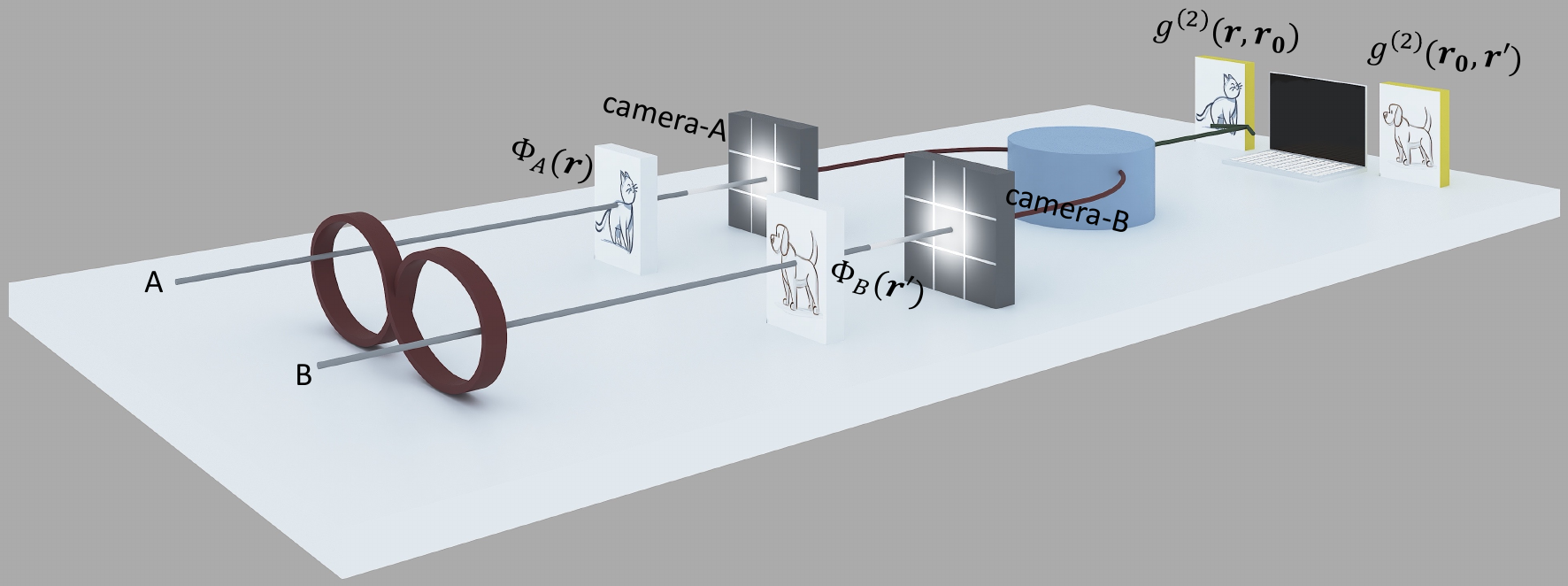}
\centering
\caption{\label{fig:1} Manipulation of quantum coherence of entangled photon pairs. Spatial light modulators (SLMs) are utilized to impose transverse-plane phases $\Phi_A(\boldsymbol{r})$ and $\Phi_B(\boldsymbol{r}')$ on the two photons, respectively. These two phases modify the second-order coherence function $g^{(2)}(\boldsymbol{r},\boldsymbol{r}')$ of photon pairs while leaving their photon-number density distributions unchanged.}
\end{figure*}

\section{Control quantum coherence via spatially varying phases \label{sec2}}
Paraxial photon pairs generated through SPDC have been routinely used in quantum sensing and quantum imaging~\cite{moreau2019imaging,gilaberte2019perspectives,taylor2013biological,Parniak2018beating}. Two paraxial photons propagating in different directions are spatially distinguishable. For convenience, we can employ two coordinate frames with path labels $A$ and $B$ [See Fig.~\ref{fig:1} (a)] to expand the WPF of each pulse in its respective frame~\cite{walborn2003multimode,deng2006spatial,toppel2012all}
\begin{align}
\left|P_{\xi}\right\rangle & =\sum_{\lambda \lambda^{\prime}} \int d\boldsymbol{k}_A\int d\boldsymbol{k}^{\prime}_B\tilde{\xi}_{\lambda \lambda^{\prime}}(\boldsymbol{k}_A,\boldsymbol{k}^{\prime}_B)\hat{a}_{\boldsymbol{k}_{A},\lambda}^{\dagger}(t)\hat{b}_{\boldsymbol{k}_{B}^{\prime},\lambda^{\prime}}^{\dagger}(t)\left|0\right\rangle ,\nonumber\\
& = \sum_{\lambda \lambda^{\prime}} \!\!\int \!\!d\boldsymbol{r}_A\!\!\int\!\! d\boldsymbol{r}^{\prime}_B\xi_{\lambda \lambda^{\prime}}(\boldsymbol{r}_A,\boldsymbol{r}^{\prime}_B,t)\hat{\psi}_{a,\lambda}^{\dagger}(\boldsymbol{r}_{A})\hat{\psi}_{b,\lambda^{\prime}}^{\dagger}(\boldsymbol{r}_{B}^{\prime})\left|0\right\rangle. \label{eq:P_RS}   
\end{align} 
where the wave-packet function (WPS) $\xi_{\lambda \lambda^{\prime}}(\boldsymbol{r}_A,\boldsymbol{r}^{\prime}_B,t)$ is the Fourier transformation of the spectrum amplitude function (SAF) $\tilde{\xi}_{\lambda \lambda^{\prime}}(\boldsymbol{k}_A,\boldsymbol{k}^{\prime}_B)$ in wave-vector space~\cite{cui2023quantum}, $\boldsymbol{r}_A$ ($\boldsymbol{k}_A$) and $\boldsymbol{r}_B$ ($\boldsymbol{k}_B$) are the coordinates (wave vectors) of the two photon pulses, and $\lambda$ is the polarization index. Usually, cylindrical coordinates $\boldsymbol{r}=\boldsymbol{\rho}+z\boldsymbol{e}_z=\rho\cos\varphi\boldsymbol{e}_x+\rho\sin\varphi\boldsymbol{e}_y+z\boldsymbol{e}_z$ ($\boldsymbol{e}_i$ is the unit vector of $i$-axis) and $\boldsymbol{k}=\tilde{\boldsymbol{\rho}}+k_z\boldsymbol{e}_z=\tilde{\rho}\cos\tilde{\varphi}\boldsymbol{e}_x+\tilde{\rho}\sin\tilde{\varphi}\boldsymbol{e}_y+k_z\boldsymbol{e}_z$  are employed to expand the WPF and SAF of paraxial photons. The two paraxial pulses can be approximately treated as two spatially independent modes. Consequently, the ladder (field) operators of two photons commute with each other~\cite{cui2023quantum}, i.e., $[\hat{a}_{\boldsymbol{k}_A,\lambda},\hat{b}_{\boldsymbol{k}_B,\lambda'}^{\dagger}] = 0$ and $[\hat{\psi}_{a,\lambda}(\boldsymbol{r}_A),\hat{\psi}_{b,\lambda'}^{\dagger}(\boldsymbol{r}_B^{\prime})] = 0$. In the following, we will omit the path labels $A$ and $B$ in the coordinates and wave vectors for conciseness.

Usually, the time-varying $g^{(2)}(\tau)$ function is used to characterize the bunching and anti-bunching property of a quasi-one-dimensional photon pulses propagating in $z$-direction. To characterize the spatial correlations of structured photon pairs, we focus on the
second-order coherence function for paraxial light pulses~\cite{yang2022quantum,Glauber1963}
\begin{equation}
g^{(2)}_{\lambda\lambda'}(\boldsymbol{r},\boldsymbol{r}')=\frac{G^{(2)}_{\lambda\lambda'}(\boldsymbol{r},\boldsymbol{r}')}{\langle\hat{n}_{a,\lambda}(\boldsymbol{r})\rangle\langle\hat{ n}_{b,\lambda'}(\boldsymbol{r}')\rangle},\label{eq:g2}
\end{equation}
where $G^{(2)}_{\lambda\lambda'}(\boldsymbol{r},\boldsymbol{r}') =\left\langle\hat{\psi}_{a,\lambda}^{\dagger}(\boldsymbol{r})\hat{\psi}_{b,\lambda'}^{\dagger}(\boldsymbol{r}')\hat{\psi}_{b,\lambda'}(\boldsymbol{r}')\hat{\psi}_{a,\lambda}(\boldsymbol{r})\right\rangle$ is the equal-time two-point intensity correlation function and the photon number densities $\langle\hat{n}_{a,\lambda}(\boldsymbol{r})\rangle=\langle\hat{\psi}^{\dagger}_{a,\lambda}(\boldsymbol{r})\hat{\psi}_{a,\lambda}(\boldsymbol{r})\rangle$ and $\langle\hat{n}_{b,\lambda'}(\boldsymbol{r}')\rangle=\langle\hat{\psi}^{\dagger}_{b,\lambda'}(\boldsymbol{r}')\hat{\psi}_{b,\lambda'}(\boldsymbol{r}')\rangle$ of the two photons. Prior research has shown that the $g^{(2)}$-function of entangled vortex photon pairs undergoes modulation due to their helical phases~\cite{yang2021quantum}. Here, we present a comprehensive approach for manipulating the spatial structure of photonic quantum coherence as shown in Fig.~\ref{fig:1}.

\subsection{Momentum-correlation-removed photon pairs \label{ses:CoherenceContro2}}
Usually, two photons generated from SPDC are correlated in both frequency and momentum due to the energy conservation and phase-matching conditions. Using a narrow-bandwidth filter or a single-mode fiber, these correlations can be removed without compromising their polarization entanglement. We start with a polarization-entangled photon pair with WPF $\xi_{\lambda\lambda'}(\boldsymbol{r},\boldsymbol{r}^\prime) =  \eta_A(\boldsymbol{r}) \eta_B(\boldsymbol{r}')\Theta_{\lambda\lambda'}$, where $\eta_A(\boldsymbol{r})$ [$\eta_B(\boldsymbol{r}')$] describes the shape of the photon-A (-B) and the $2\times 2$ matrix $\Theta_{\lambda\lambda'}$ characterize the polarization state. Without loss of generality, we consider the polarization-entangled state
\begin{equation}
\Theta_{\lambda\lambda'} =\frac{1}{\sqrt{2}}(\delta_{\lambda,H}\delta_{\lambda',V} + \delta_{\lambda,V}\delta_{\lambda',H}).    
\end{equation}
Our method can be directly applied to other entangled photon pairs. To modulate the quantum coherence function, two polarization-sensitive SLMs are utilized to impose distinct phase patterns onto the two photons
\begin{equation}
\!\!\!\xi_{\lambda\lambda'}(\boldsymbol{r},\boldsymbol{r}^\prime) \! = \!  \frac{\eta_A(\boldsymbol{r}) \eta_B(\boldsymbol{r}')}{\sqrt{2}}\! \left[\delta_{\lambda,H}\delta_{\lambda',V} e^{i\Phi_A (\boldsymbol{r})} \!+\!\delta_{\lambda,V}\delta_{\lambda',H}e^{i\Phi_B (\boldsymbol{r}')}\!\right]\!.
\end{equation}
Here, spatially varying phases $\Phi_{A}(\boldsymbol{r})$ and $\Phi_{B}(\boldsymbol{r}')$ have only been added to the $|H\rangle$-state photons. Usually, a SLM only changes the phase of photons within their transverse plane. However, our approach can apply to cases involving three-dimensional structured phases as well.

\begin{figure}
\includegraphics[width=8cm]{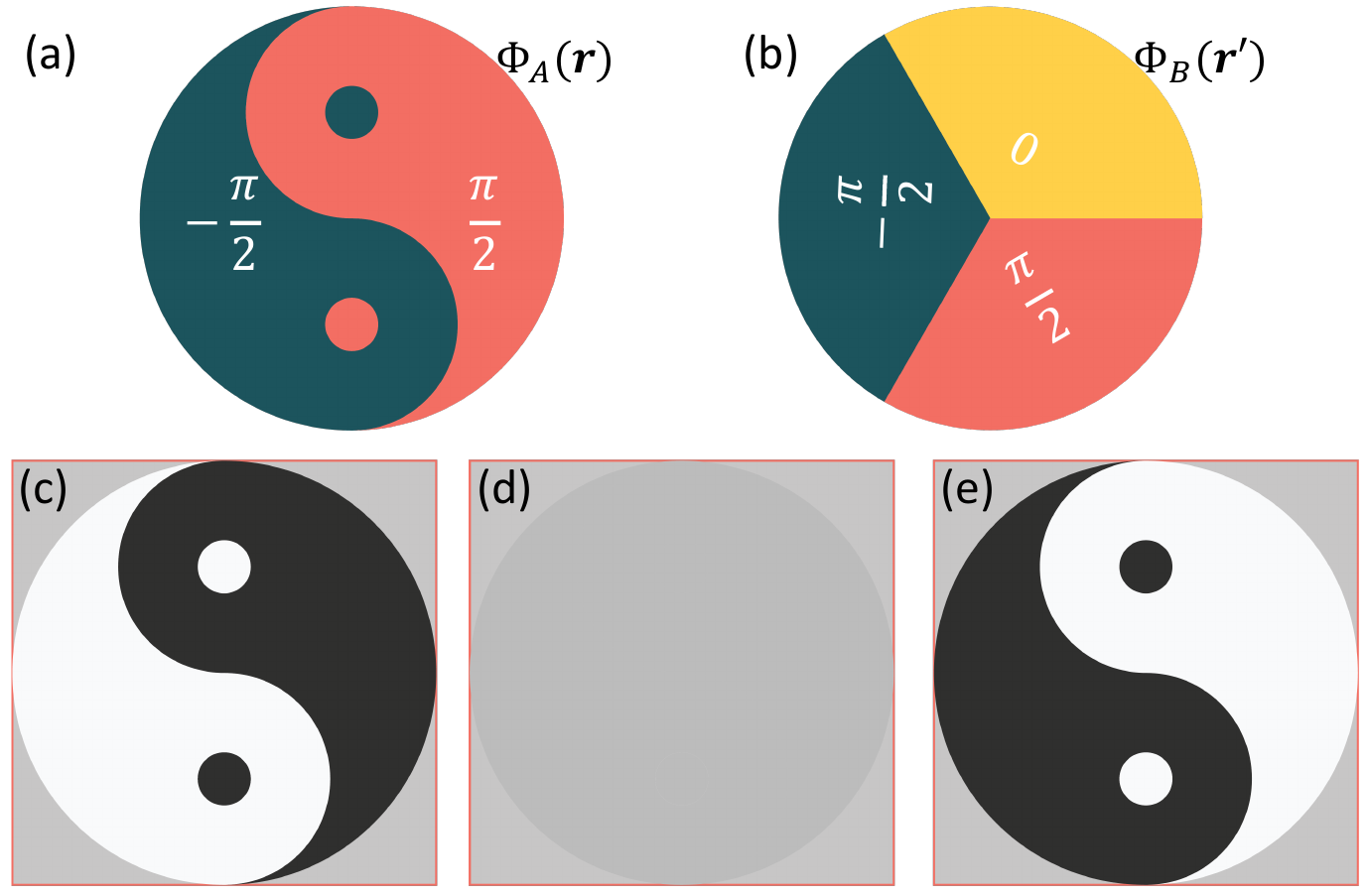}
\caption{\label{fig:2} (a) and (b) are phase patterns applied on photon-A and photon-B, respectively. (c-e) are coincidence images obtained from photon-A when $\boldsymbol{r}'$ of photon-B is anchored in the regions with $\Phi_B(\boldsymbol{r}')=\pi/2$, $\Phi_B(\boldsymbol{r}')=0$, and $\Phi_B(\boldsymbol{r}')=-\pi/2$, respectively. 
}
\end{figure}

The added phases do not change the field intensity distributions of the two photons, i.e., $\langle \hat{n}_{a,\lambda}(\boldsymbol{r})\rangle=|\eta_A(\boldsymbol{r})|^2/2$ and $\langle \hat{n}_{b,\lambda}(\boldsymbol{r}')\rangle =|\eta_B(\boldsymbol{r}')|^2/2$. 
Furthermore, we can confirm that the applied phase cannot be extracted using polarization filters, such as circular-polarization projections
\begin{align}
\langle\hat{\psi}^{\dagger}_{a,L}(\boldsymbol{r})\hat{\psi}_{a,L}(\boldsymbol{r})\rangle & =\langle\hat{\psi}^{\dagger}_{a,R}(\boldsymbol{r})\hat{\psi}_{a,R}(\boldsymbol{r})\rangle= \frac{1}{2}|\eta_A(\boldsymbol{r})|^2, \\
\langle\hat{\psi}^{\dagger}_{b,L}(\boldsymbol{r}')\hat{\psi}_{b,L}(\boldsymbol{r}')\rangle & =\langle\hat{\psi}^{\dagger}_{b,R}(\boldsymbol{r}')\hat{\psi}_{b,R}(\boldsymbol{r}')\rangle= \frac{1}{2}|\eta_B(\boldsymbol{r}')|^2,
\end{align}
with field operators $\hat{\psi}_L(\boldsymbol{r})=\left[\hat{\psi}_H(\boldsymbol{r})+i\hat{\psi}_V(\boldsymbol{r})\right]/\sqrt{2}$ and $\hat{\psi}_R(\boldsymbol{r})=\left[\hat{\psi}_H(\boldsymbol{r})-i\hat{\psi}_V(\boldsymbol{r})\right]/\sqrt{2}$.
Finally, we measure the second-order coherence function associated with the correlation between two left-circular-polarization states
\begin{equation}
g^{(2)}_{LL}(\boldsymbol{r},\boldsymbol{r}') =  1+ \cos[\Phi_A(\boldsymbol{r})+\Phi_B(\boldsymbol{r}')].\label{eq:g2LL}  
\end{equation}
By manipulating the phases $\Phi_A(\boldsymbol{r})$ and $\Phi_B(\boldsymbol{r}')$, we can tailor the spatial structure of the quantum coherence of a photon pair for purpose, such as incorporating a dog onto photon-$A$ and a cat onto photon-$B$ (see Fig.~\ref{fig:1}). A similar coherence function can be obtained for right-circularly polarized photons. 

To extract the image encoded in the phases, we can anchor one of the coordinates (e.g., by setting $\boldsymbol{r}'=\boldsymbol{r}_0$) within the coherence function $g^{(2)}_{LL}(\boldsymbol{r},\boldsymbol{r}_0)$ and scan the other coordinate.  As illustrated in Fig.~\ref{fig:2}, we imprint a Taiji pattern [panel (a)] on photon-A and a three-sector pattern [panel (b)] on photon-B. For the fixing $\boldsymbol{r}'$ in the region where $\Phi_B (\boldsymbol{r}')=\pi/2$, the pattern in panel (c) will be observed in coincidence imaging. Similarly, fixing $\boldsymbol{r}'$ in the regions where $\Phi_B (\boldsymbol{r}')=0$ and $\Phi_B (\boldsymbol{r}')=-\pi/2$, the patterns depicted in panels (d) and (e), respectively, will be obtained from the measured coherence functions.

\begin{figure}
\includegraphics[width=8.5cm]{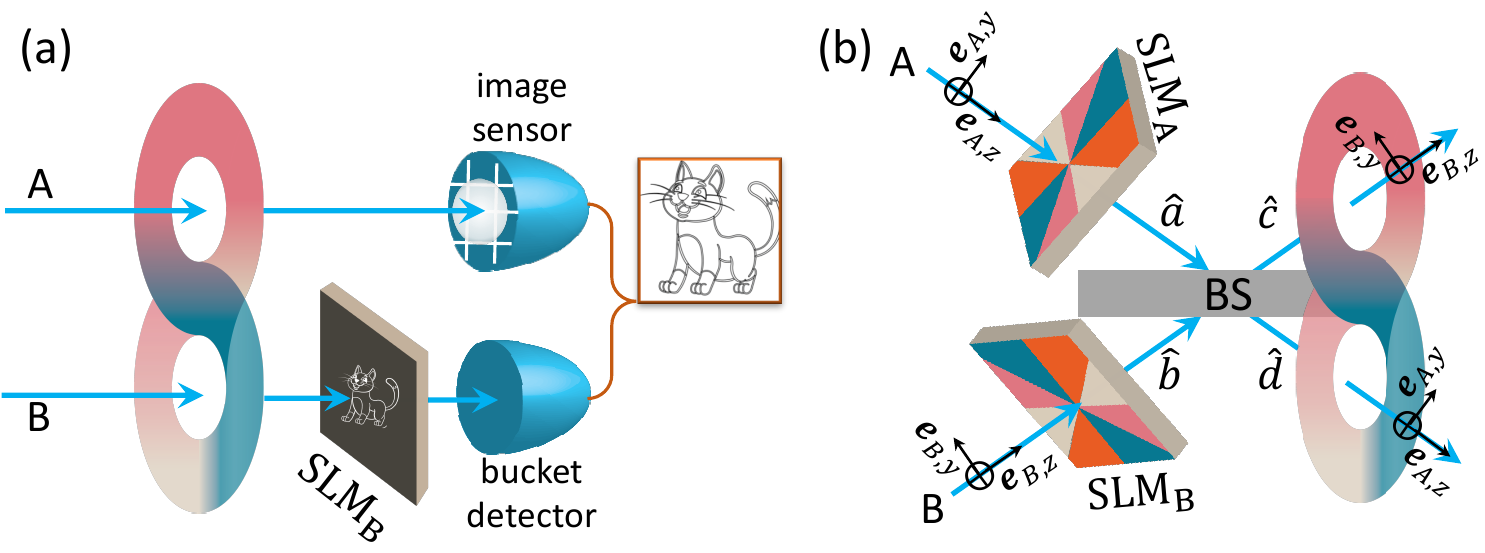}
\centering
\caption{\label{fig:3} (b) Controlling the quantum coherence of strongly correlated pairs. The pattern applied on the phase of photon-B can be extracted from coincidence imaging from an imaging sensor in path-A. (b) Controlling the quantum coherence of product-state photon pairs with Hong-Ou-Mandel (HOM) interference. }
\end{figure}

\subsection{Strongly correlated photon pairs\label{ses:CoherenceContro2}}
The coherence function of photon pairs with momentum correlations is typically modified by added phases as well as the intensity distributions, unlike the simple result in Eq.~(\ref{eq:g2LL}). A uniformly polarized photon pair generated through SPDC processes can be generally described by a SAF~\cite{Hong1985SPDC,Monken1998SPDC,Black2019nonlocal}
$\tilde{\xi}_{\lambda\lambda'}(\boldsymbol{k},\boldsymbol{k}') = \tilde{\eta}(k_z,k'_z)\tilde{\mathcal{E}}(\tilde{\boldsymbol{\rho}}+\tilde{\boldsymbol{\rho}}')\tilde{\chi}(\tilde{\boldsymbol{\rho}}-\tilde{\boldsymbol{\rho}}')\Theta_{\lambda,\lambda'}$. Here, the function $\tilde{\eta}(k_z,k'_z)$ determines the shapes of the two photons along their propagating directions, while $\tilde{\mathcal{E}}(\tilde{\boldsymbol{\rho}}+\tilde{\boldsymbol{\rho}}')$ characterizes the angular profile of the pump beam in the transverse plane. The phase-matching condition can be described by
\begin{equation}
\tilde{\chi}(\tilde{\boldsymbol{\rho}}-\tilde{\boldsymbol{\rho}}')=\frac{1}{\pi}\sqrt{\frac{2L}{k_0}}{\rm sinc}\left(\frac{L|\tilde{\boldsymbol{\rho}}-\tilde{\boldsymbol{\rho}}'|^{2}}{4k_0}\right),   
\end{equation}
where $k_0$ is the wavenumber of the pump and $L$ is the thickness of the nonlinear crystal. For a pump with a flat transverse distribution, the function $\tilde{\mathcal{E}}(\tilde{\boldsymbol{\rho}}+\tilde{\boldsymbol{\rho}}')$ approaches to a Dirac delta function $\delta (\tilde{\boldsymbol{\rho}}+\tilde{\boldsymbol{\rho}}')$. While for a thin crystal, the $\rm{sinc}$ function is relatively flat in the momentum space and its Fourier transformation could be approximated as a delta function $\chi(\boldsymbol{\rho}-\boldsymbol{\rho}')\approx\delta(\boldsymbol{\rho}-\boldsymbol{\rho}')$~
\cite{Camphausen2021quantum,Black23quantum}. In this case, two photons are strongly correlated in momentum, as well as in real space.

We illuminate a particular case, in which a structured phase $\Phi_B (\boldsymbol{\rho}')$ is only applied to photon-B as shown in Fig.~\ref{fig:3} (a). The WPF function of the photon pair after the SLM can be obtained through Fourier transformation of the SAF
\begin{align}
\xi_{\lambda\lambda'}(\boldsymbol{r},\boldsymbol{r}^\prime) = & \frac{1}{\sqrt{2}}\eta(z,z')\mathcal{E}(\boldsymbol{\rho}\!+\!\boldsymbol{\rho}')\chi (\boldsymbol{\rho}\!-\!\boldsymbol{\rho}') \nonumber\\
& \times \left[\delta_{\lambda,H}\delta_{\lambda',V} + \delta_{\lambda,V}\delta_{\lambda',H}e^{i\Phi_B (\boldsymbol{\rho}')}\right].
\end{align}
In the absence of momentum correlations, the structured phase on photon-B can solely be extracted by the image sensor in path-B. However, for a strongly correlated photon pair with $\chi(\boldsymbol{\rho}-\boldsymbol{\rho}')\approx\delta(\boldsymbol{\rho}-\boldsymbol{\rho}')$, the phase $\Phi_B$ can also be extracted by a camera in path-A via coincidence measurements
\begin{align}
g_{{\rm eff}}^{(2)}(\boldsymbol{\rho}) & =\frac{\int d\boldsymbol{r}'\left\langle \langle\hat{\psi}_{a,L}^{\dagger}(\boldsymbol{r})\langle\hat{\psi}_{b,L}^{\dagger}(\boldsymbol{r}')\hat{\psi}_{b,L}(\boldsymbol{r}')\rangle\hat{\psi}_{a,L}(\boldsymbol{r})\rangle\right\rangle }{\langle\hat{\psi}_{a,L}^{\dagger}(\boldsymbol{r})\hat{\psi}_{a,L}(\boldsymbol{r})\rangle}\\
 & =\frac{1}{2}\left[1+\cos\Phi_B(\boldsymbol{\rho})\right].
\end{align}
This is similar to the ghost imaging~\cite{Pittman1995}, but here we focus more on using quantum entanglement to control the quantum coherence of photon pairs.

\section{Coherence control via HOM interference\label{sec3}}
Our method primarily relies on the utilization of global entanglement to manipulate the detailed spatial structure of the quantum coherence functions. Hong-Ou-Mandel (HOM) interference is a viable approach for generating two-path entanglement~\cite{chrapkiewicz2016hologram}, enabling effective manipulation of the optical coherence even for product-state photon pairs as depicted in Fig.~\ref{fig:3} (b). When two linearly polarized photons are initially in a product state, the WPF of the resulting photon pair after the SLMs can be expressed as follows
\begin{equation}
\xi(\boldsymbol{r},\boldsymbol{r}^{\prime})=\eta_A (\boldsymbol{r})\eta_B(\boldsymbol{r}')e^{i\left[\Phi_A(\boldsymbol{r})+\Phi_B(\boldsymbol{r}')\right]}.
\end{equation}
where the real functions $\Phi_A (\boldsymbol{r})$ and $\Phi_B (\boldsymbol{r}')$ denote the phases imposed on the two photons, respectively. For simplicity, we assume that $\eta_A (\boldsymbol{r})$ and $\eta_B (\boldsymbol{r}^\prime)$ possess axial symmetry and are independent of the azimuthal angles $\varphi$ and $\varphi'$. However, we note that our approach can be extended to encompass more general scenarios as well. 

The complete theory of HOM interference for structured photon pairs encompassing spectral, polarization, and spatial degrees of freedom has been established~\cite{deng2006spatial,supplementary} and further extended to mixed-state photon pairs~\cite{toppel2012all}. Here, we show that the phases added by SLMs do not affect the photon number density distribution of photon pulse at each output channel of an HOM interferometer 
\begin{align}
\left\langle \Psi_{{\rm out}}\mid\hat{\psi}_{c}^{\dagger}(\boldsymbol{r})\hat{\psi}_{c}(\boldsymbol{r})\mid\Psi_{{\rm out}}\right\rangle & = \frac{1}{2}\left[ \left|\eta_A(\boldsymbol{r})\right|^{2}+\left|\eta_B(\boldsymbol{r})\right|^{2}\right],\\
\left\langle \Psi_{{\rm out}}\mid\hat{\psi}_{d}^{\dagger}(\boldsymbol{r})\hat{\psi}_{d}(\boldsymbol{r})\mid\Psi_{{\rm out}}\right\rangle
& = \frac{1}{2}\left[\left|\eta_A(\boldsymbol{r})\right|^{2}+\left|\eta_B(\boldsymbol{r})\right|^{2}\right],
\end{align}
where $|\Psi_{{\rm out}}\rangle$ is the output state after HOM interference~\cite{supplementary}. The terms involving the phases $\Phi_A$ and $\Phi_B$ undergo complete cancellation due to the destructive interference occurring between the output bunching and anti-bunching  photons~\cite{cui2023quantum}. However, the added two phases change the two-port $g^{(2)}$-function of output photons significantly\begin{widetext}
\begin{align}
g^{(2)}_{cd}  (\boldsymbol{r},\boldsymbol{r}') = & \frac{\left\langle\Psi_{\rm out}\mid \hat{\psi}^{\dagger}_d(\boldsymbol{r})\hat{\psi}^{\dagger}_c(\boldsymbol{r}')\hat{\psi}_c(\boldsymbol{r}')\hat{\psi}_d(\boldsymbol{r})\mid\Psi_{\rm out}\right\rangle}{ \left\langle \Psi_{{\rm out}}\mid\hat{\psi}_{d}^{\dagger}(\boldsymbol{r})\hat{\psi}_{d}(\boldsymbol{r})\mid\Psi_{{\rm out}}\right\rangle \left\langle \Psi_{{\rm out}}\mid\hat{\psi}_{c}^{\dagger}(\boldsymbol{r}')\hat{\psi}_{c}(\boldsymbol{r}')\mid\Psi_{{\rm out}}\right\rangle} \\
= & \frac{|\eta_A (\boldsymbol{r})|^2|\eta_B (\boldsymbol{r}')|^2+|\eta_B (\boldsymbol{r})|^2|\eta_A (\boldsymbol{r}')|^2-\left\{\eta^*_A(\boldsymbol{r})\eta_B(\boldsymbol{r})\eta^*_B(\boldsymbol{r}')\eta_A(\boldsymbol{r}')e^{-i[\Phi_A(\boldsymbol{r})-\Phi_B(\bar{\boldsymbol{r}})+\Phi_B(\boldsymbol{r}')-\Phi_A(\bar{\boldsymbol{r}}')]}+{\rm c.c.}\right\}}{\left[\left|\eta_A(\boldsymbol{r})\right|^{2}+\left|\eta_B(\boldsymbol{r})\right|^{2}\right]\left[ \left|\eta_A(\boldsymbol{r}')\right|^{2}+\left|\eta_B(\boldsymbol{r}')\right|^{2}\right]}, \label{eq:gcd1}
\end{align}
\end{widetext}
where $\bar{\boldsymbol{r}}=\{x,-y,z\}$ characterizes the influence of the reflection operation as shown in Fig~\ref{fig:3} (b).
For two input photons of the same pulse shape (i.e., $|\eta_A(\boldsymbol{r})|=|\eta_B (\boldsymbol{r})|$), the $g^{(2)}$-function reduces to 
\begin{equation}
 g^{(2)}_{cd} (\boldsymbol{r},\boldsymbol{r}') \!=\! \frac{1}{2}\left\{1\!-\!\cos [\Phi_A(\boldsymbol{r})\!-\!\Phi_B(\bar{\boldsymbol{r}})\!+\!\Phi_B(\boldsymbol{r}')\!-\!\Phi_A(\bar{\boldsymbol{r}}')]\right\}. \label{eq:g2cd}  
\end{equation}
The $g^{(2)}$-function of the input product-state photon pairs is a constant $1$. The modulation observed in the coherence function, induced by the imposed phases, is entirely attributed to the two-body entanglement generated through HOM interference.

\subsection{Helical continuous phases}
The helical phase structure of twisted photons carrying non-vanishing OAM can modify the second-order optical coherence function~\cite{yang2022quantum} and lead to interesting HOM interference effects~\cite{zhang2016engineering,Ambrosio2019tunable}. HOM interference of two photons hyperentangled between spin and OAM degrees has also been experimentally demonstrated~\cite{liu2022hong}. Here, we consider a scenario in which helical phases $\Phi_A (\boldsymbol{r})= \Phi_B (\boldsymbol{r}) = m \varphi$ are introduced into the two input channels. No HOM dip or peak will be observed except for $m=0$~\cite{cui2023quantum,supplementary}. However, the azimuthal angles modulate the coherence function $g^{(2)} (\boldsymbol{r},\boldsymbol{r}') = [1 - \cos (2m\varphi+2m\varphi')]/2$ of the output photons. Our method could also be applied to cases where the OAM quantum numbers of the two input channels are different.

In the experiment~\cite{zhang2016engineering}, Zhang et al. demonstrate the HOM dip of entangled twisted photon pairs. By tuning the orientation angle difference $\varphi_0$ of a pair of Dove prisms, they added a relative phase ($|m\rangle\rightarrow |m\rangle \exp(i2m\varphi_0)$) between the two entangled photons
\begin{equation}
\xi(\boldsymbol{r},\boldsymbol{r}',t) =\mathcal{N}\eta_m (\boldsymbol{r},t)\eta_{m}(\boldsymbol{r}',t) \left[e^{im(\varphi+\varphi^{\prime})} e^{i 2 m \varphi_0}+ {\rm c.c.}\right]\!,
\end{equation}
which can be re-expressed as a superposition of exchange-reflection symmetric and anti-symmetric states. The two-port coincidence probability $P^{(2)}_{cd} = [1 - \cos (4m\varphi_0)]/2$ varies with the angle $\phi_0$ and an HOM dip gradually turns to a peak. The entangled twisted photon pair can be exploited for quantum sensing of the rotational angle $\varphi_0$~\cite{maga2014amplification} with sensitivity $\propto 2m$. Here, we show that the two-port $g^{(2)}$-function is additionally modulated by the angle $\varphi_0$ as well as the two azimuthal angles $\varphi$ and $\varphi'$,
\begin{equation}
  g^{(2)}_{cd} (\boldsymbol{r},\boldsymbol{r}') = \frac{1}{2}\{1-\cos [2m(\varphi+\varphi')]\}[1-\cos(4m\varphi_0)].  
\end{equation}

\begin{figure}
\includegraphics[width=8.5cm]{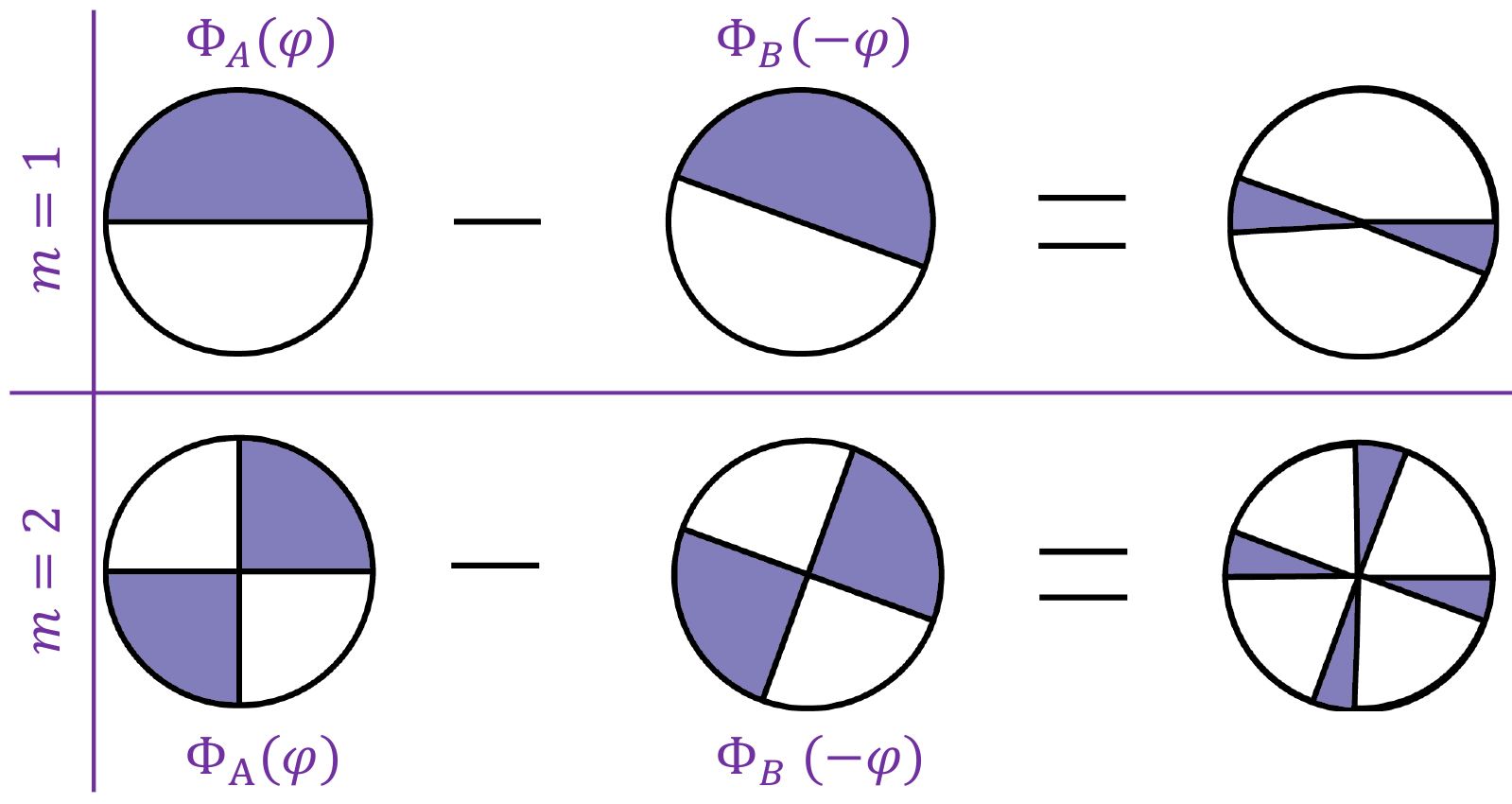}
\centering
\caption{\label{fig:4} Phase difference $\Phi_A(\varphi)-\Phi_B(-\varphi)$. (a) The transverse plane is split into two half-circular discs. (b) The transverse plane is split into four equal sectors. The phase of dark regions is $\pi$ or $-\pi$. The phase of white regions is $0$.}
\end{figure}

\subsection{Discontinuous circular-sector phases}
The spatially varying phase of photons can provide more powerful ways to control photonic quantum coherence in addition to the helical structure of twisted light. Using a spatial light modulator or a metasurface~\cite{Yu2011Light,devlin2017Arbitrary}, we can imprint arbitrary patterns on the two input photons as shown in previous section. Here, we consider the simplest case where the transverse plane is split into $2m$ ($m$ is an integer) same-sized circular sectors (see Fig.~\ref{fig:4}). 

The spatially varying phases are given by
\begin{equation}
\Phi_A(\varphi)=\begin{cases}
\pi, & \varphi\in[(2j-2)\pi/m,(2j-1)\pi/m)\\
0, & \varphi\in[(2j-1)\pi/m,2j\pi/m)
\end{cases},\label{eq:Phi}
\end{equation}
and 
\begin{equation}
\Phi_B(\varphi)=\begin{cases}
0, & \varphi\in[(2j-2)\pi/m+\varphi_{0},(2j-1)\pi/m+\varphi_{0})\\
\pi, & \varphi\in[(2j-1)\pi/m+\varphi_{0},2j\pi/m+\varphi_{0})
\end{cases}, \label{eq:Phiprime}
\end{equation}
where the integer  $j$ runs from $1$ to $m$ and $\varphi_0$ is a mismatch angle of the two phases. The HOM coincidence probability for $\tau = 0$ and $\varphi_0 \in [0,\pi/m)$ is given by
\begin{align}
P_{cd}^{(2)} = & \frac{1}{2}\!-\!\frac{1}{4}\!\int_0^{2\pi}\!\! \frac{d\varphi}{2\pi}\!\!\int_0^{2\pi}\!\! \frac{d\varphi^{\prime}}{2\pi}\! \left\{ e^{-i\left[\Phi_A(\varphi)-\Phi_B(-\varphi)+\Phi_B(\varphi^{\prime})-\Phi_A(-\varphi^{\prime})\right]} \right. \\
& \left. +e^{-i\left[\Phi_A(-\varphi)-\Phi_B(\varphi)+\Phi_B(-\varphi^{\prime})-\Phi_A(\varphi^{\prime})\right]} \right\}\\
= & \frac{1}{2}\left[1-\left(1-\frac{m\varphi_0}{\pi}\right)^2\right],
\end{align}
where we have used the normalization condition $\int_0^{\infty}\rho d\rho\int_{-\infty}^{\infty} dz|\eta(\boldsymbol{r},t)|^2=1/2\pi$ and the phase difference shown in Fig.~\ref{fig:4}.

\begin{figure}
\includegraphics[width=8.5cm]{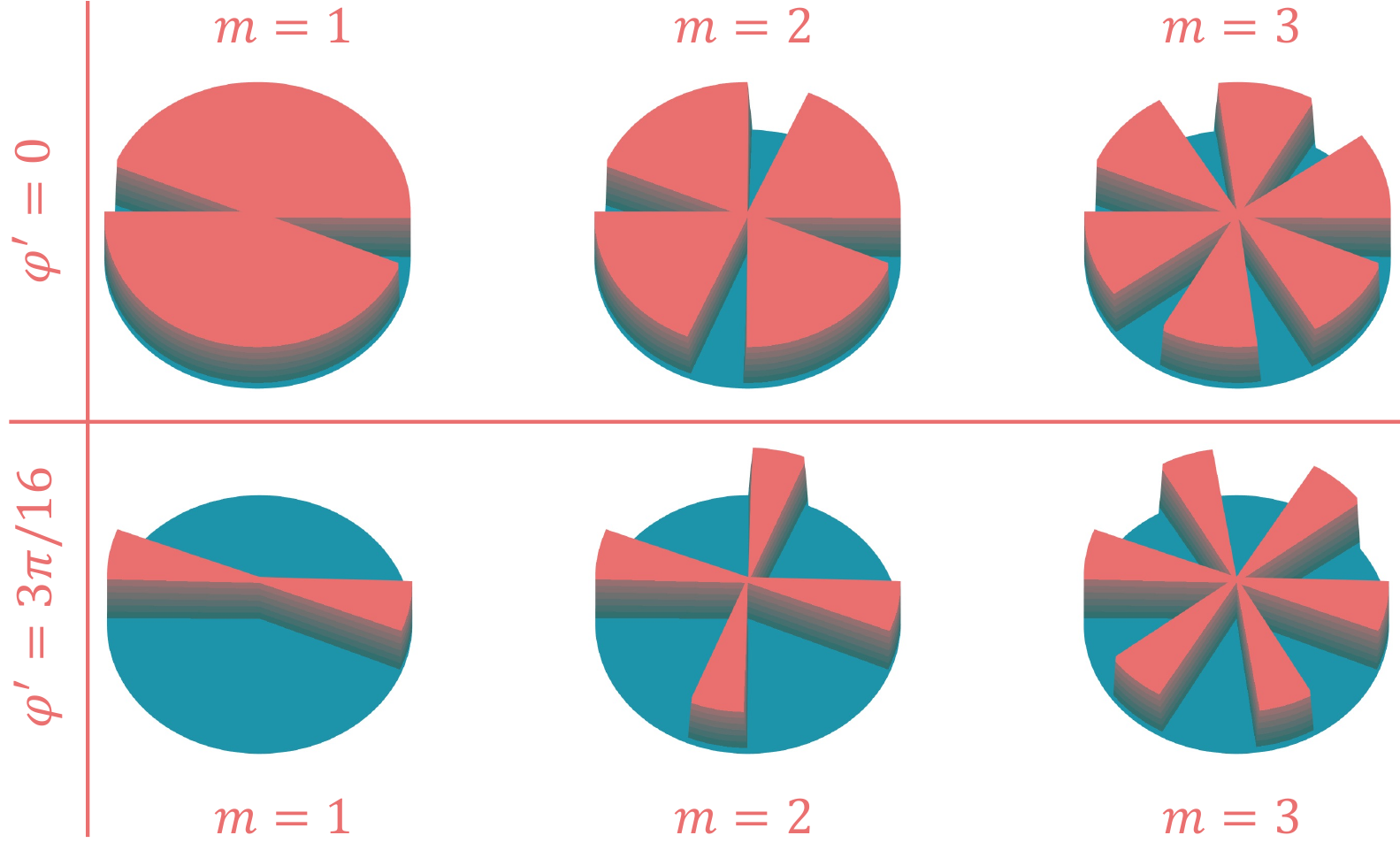}
\caption{\label{fig:5} The $g^{(2)}_{cd}(\boldsymbol{r},\boldsymbol{r}')$ function vary with the azimuthal angle $\varphi$. The transverse plane of the two photons has been split into $2m$ same-sized circular sectors with mismatch angle $\varphi_0 = \pi/8$. The azimuthal angle for the top and bottom rows are set as $\varphi'=0$ and $\varphi'=3\pi/16$, respectively. The values of the $g^{(2)}$-function in the raised and flat areas are $1$ and $0$, respectively.
}
\end{figure}

Our primary focus lies in the manipulation of the detailed coherence function through the engineering of phase profiles in the transverse plane. The photon-number densities at the two output ports are given by $n_{c}(\boldsymbol{r})=\left|\eta(\boldsymbol{r})\right|^{2}$ and $n_{d}(\boldsymbol{r})=\left|\eta(\boldsymbol{r})\right|^{2}$, respectively. Similarly, the imposed phases play a role in shaping the second-order coherence function, as described in Eq.~(\ref{eq:g2cd}). Note that $\varphi$ changes to $-\varphi$ for $\bar{\boldsymbol{r}}$. In Fig.~\ref{fig:5}, we showcase the $g^{(2)}$-function while varying $\varphi$ and and keeping $\varphi'$ fixed. The mismatch angle is set as $\varphi_0 = \pi/8$. In the top row, $\boldsymbol{r}'$ is pinned at $\varphi'=0$. The $g^{(2)}$-function vanishes in regions where $\Phi_A(\varphi)-\Phi_B(-\varphi)=\pm \pi$ as shown in Fig.~\ref{fig:4}, given that $\Phi_B(\varphi')=0$ and $\Phi_A(\varphi')=\pi$. In the bottom row, we let $\varphi' = 3\pi/16$. Consequently, the $g^{(2)}$-function vanishes in the complementary regions since $\Phi_B(\varphi')=\Phi_A(-\varphi')=0$.

\begin{figure}
\includegraphics[width=7cm]{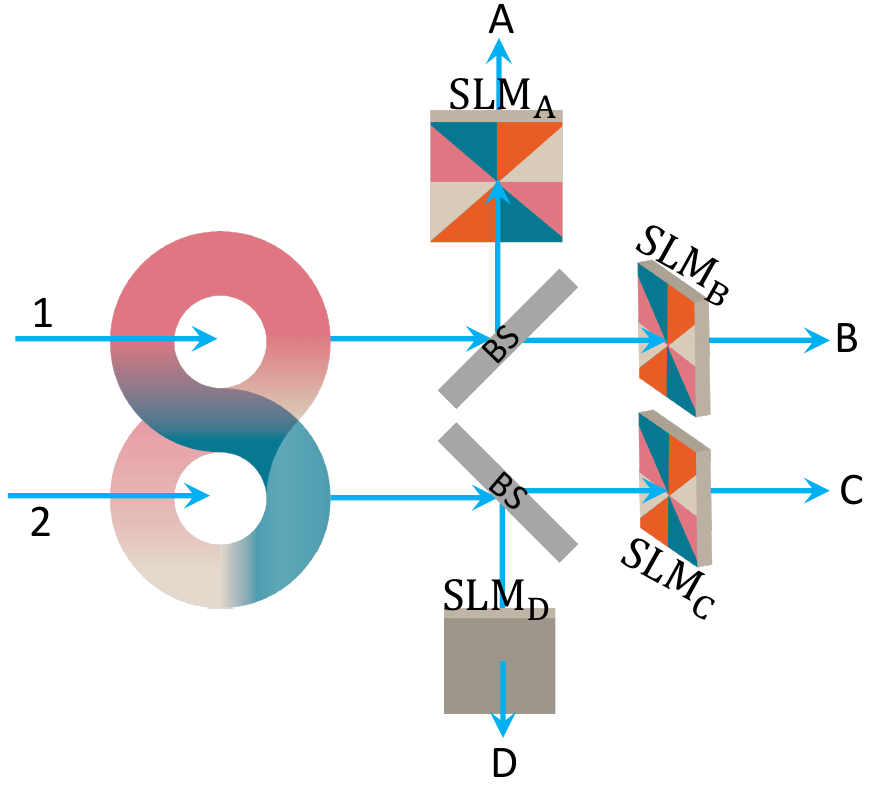}
\caption{\label{fig:6} Fourth-order coherence control scheme. The four photons are initially prepared in an entangled state $~(|HH\rangle_1 |VV\rangle_2+|VV\rangle_1 |HH\rangle_2)$. To generate four output ports, two beam splitters (BSs) are employed. Four spatial light modulators (SLMs) impose four different phase patterns on the four photons.
}
\end{figure}

\begin{figure}
\vspace{0.5cm}
\includegraphics[width=7.5cm]{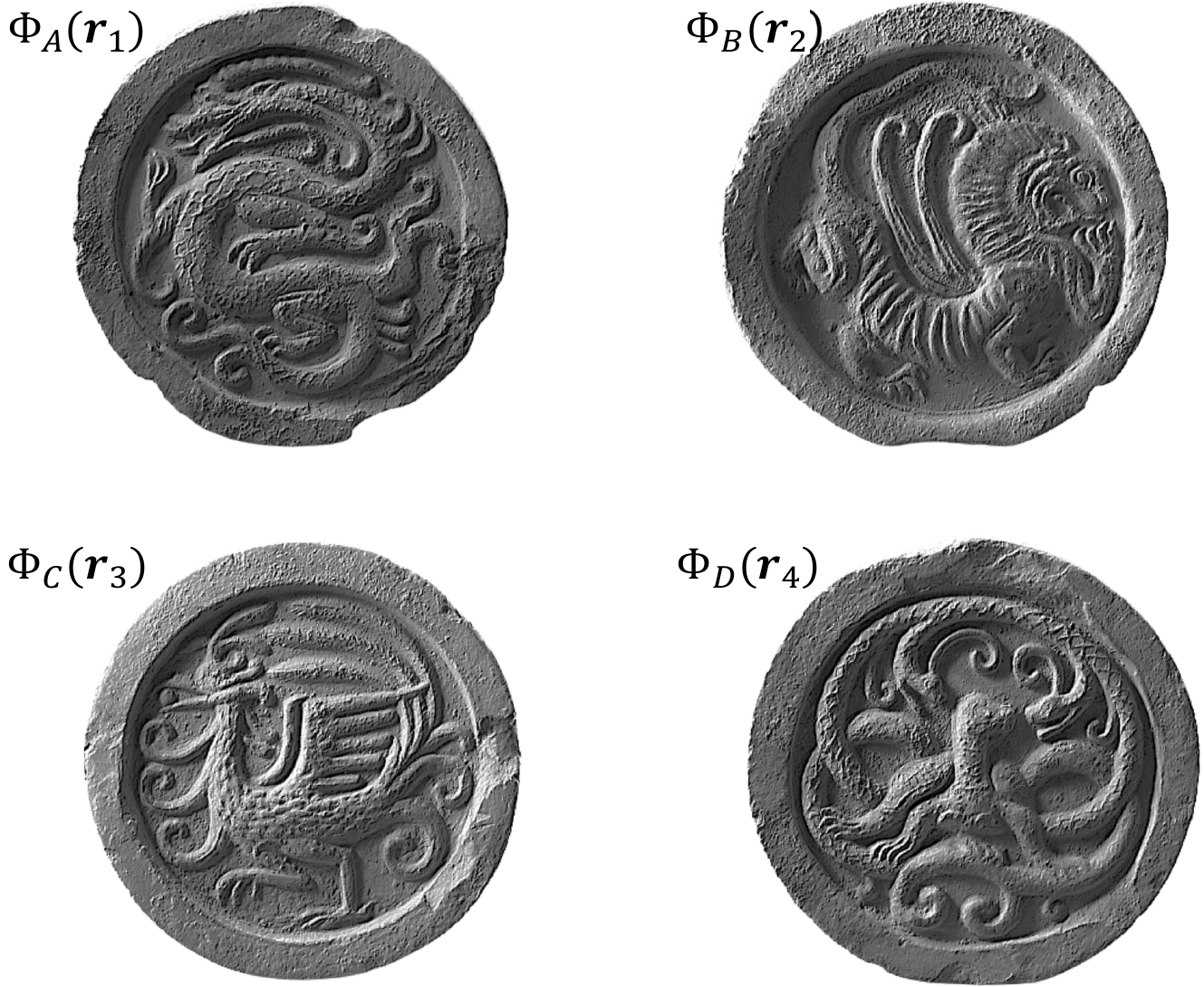}
\caption{\label{fig:7} Four Chinese mythological creatures corresponding to the four quadratures of I Ching could be imposed on the phases of the four photons. They are the Azure Dragon of the East [$\Phi_A (\boldsymbol{r}_1)$],  the White Tiger of the West [$\Phi_B (\boldsymbol{r}_2)$],  the Vermilion Bird of the South [$\Phi_C (\boldsymbol{r}_3)$], and the Black Tortoise of the North [$\Phi_D (\boldsymbol{r}_4)$]. The photos are from the \href{https://www.qzhwbwg.com/sswd}{Xi'an Qin Brick and Han Tile Museum}.
}
\end{figure}

Discontinuous jump exists in our simulation of the $g^{(2)}$-function in Fig.~\ref{fig:5}. However, in experimental settings, a continuous and sharp change would be observed instead. This sharp change presents an opportunity for utilizing the coherence function as a tool for high-sensitivity quantum sensing of the rotational angle $\varphi_0$. By combining this effect with the geometric rotation of photons traveling through a coiled fiber~\cite{Yang2023Geometric,alexeyev2006topological,Alexeyev2006}, the sharp change in the coherence function of photon pairs can be harnessed for the exploration of a novel type of laser gyroscope, similar to those based on the Sagnac effect~\cite{Chow1985Sagnac}.

\section{Higher-order coherence control\label{sec4}}
Our method can also be extended to control higher-order coherence of photons. Here, we only illustrate its application using four-photon pulses as an example as depicted in Fig.~\ref{fig:6}. We begin with a path-entangled four-photon state $\sim (|HH\rangle_1 \otimes |VV\rangle_2 + |VV\rangle_1 \otimes |HH\rangle_2/\sqrt{2}$). For simplicity, let us consider the case where the four photons share the same shape, characterized by the wavepacket function (WPF) $\eta (\boldsymbol{r})$, such as a Gaussian profile. Following the two beam splitters, the state of the four photons can be described as\begin{widetext}
 \begin{align}
 |\Psi\rangle & =\frac{1}{8\sqrt{2}}\int d\boldsymbol{r}_{1}\int d\boldsymbol{r}_{2}\int d\boldsymbol{r}_{3}\int d\boldsymbol{r}_{4}\eta(\boldsymbol{r}_{1})\eta(\boldsymbol{r}_{2})\eta(\boldsymbol{r}_{3})\eta(\boldsymbol{r}_{4})\nonumber \\
 & \left\{ \left[i\hat{\psi}_{A,H}^{\dagger}(\boldsymbol{r}_{1})+\hat{\psi}_{B,H}^{\dagger}(\boldsymbol{r}_{1})\right]\left[i\hat{\psi}_{A,H}^{\dagger}(\boldsymbol{r}_{2})+\hat{\psi}_{B,H}^{\dagger}(\boldsymbol{r}_{2})\right]\left[\hat{\psi}_{C,V}^{\dagger}(\boldsymbol{r}_{3})+i\hat{\psi}_{D,V}^{\dagger}(\boldsymbol{r}_{3})\right]\left[\hat{\psi}_{C,V}^{\dagger}(\boldsymbol{r}_{4})+i\hat{\psi}_{D,V}^{\dagger}(\boldsymbol{r}_{4})\right]+(H\leftrightarrow V)\right\} \left|0\right\rangle, 
 \label{eq:State4photon}
 \end{align}   
\end{widetext}
where the WPF is expressed in four different coordinate frames determined by the propagating axes of the photons and we have used the fact that $\eta(\boldsymbol{r})$ possesses reflection symmetry, $\eta (\bar{\boldsymbol{r}})=\eta(\boldsymbol{r})$. The field operators satisfy 
 the commutation relation $[\psi_{i,\lambda}(\boldsymbol{r}),\hat{\psi}^{\dagger}_{j,\lambda'}(\boldsymbol{r}')]=\delta_{ij}\delta_{\lambda\lambda'}\delta (\boldsymbol{r}-\boldsymbol{r}')$.

Next, we utilize four SLMs to apply distinct phase patterns to the $H$-polarized photons in the four output channels. For example, we can imprint the four Chinese mythological creatures [as depicted in Fig.~\ref{fig:7}] representing the four quadratures of I Ching on the photons. The final quantum state $\left|\tilde{\Psi}\right\rangle$ of the four photons is obtained by substituting the $H$-polarization operator $\psi^{\dagger}_{i,H}(\boldsymbol{r})$ ($i=\{A,B,C,D\}$) in Eq.~(\ref{eq:State4photon}) with $\psi^{\dagger}_{i,H}(\boldsymbol{r})\exp[i\Phi_i (\boldsymbol{r})]$.

The imposed four phases will not change the distribution of photon number density at each output port
\begin{equation}
 n_i (\boldsymbol{r}) = \sum_{\lambda}\left\langle\tilde{\Psi}\right|\hat{\psi}^{\dagger}_{i,\lambda}\hat{\psi}_{i,\lambda}\left|\tilde{\Psi}\right\rangle =  |\eta(\boldsymbol{r}_i)|^2.  
\end{equation}
We can also evaluate the second and higher order correlations of the four-photon state\begin{widetext}
\begin{align}
 G^{(2)}_{ij}(\boldsymbol{r}_{1},\boldsymbol{r}_{2}) & =\sum_{\lambda_{1}\lambda_{2}}\left\langle \tilde{\Psi} \right|\hat{\psi}_{i,\lambda_{1}}^{\dagger}(\boldsymbol{r}_{1})\hat{\psi}_{j,\lambda_{2}}^{\dagger}(\boldsymbol{r}_{2})\hat{\psi}_{j,\lambda_{2}}(\boldsymbol{r}_{2}) \hat{\psi}_{i,\lambda_{1}}(\boldsymbol{r}_{1}) \left|\tilde{\Psi} \right\rangle ,\\
 & =\frac{1}{2}|\eta(\boldsymbol{r}_{1})|^{2}|\eta(\boldsymbol{r}_{2})|^{2}\left[\delta_{ij}+\delta_{i,A}\delta_{j,B}+\delta_{i,C}\delta_{j,D}+2(\delta_{i,A}+\delta_{i,B})(\delta_{j,C}+\delta_{j,D})\right],
\end{align}
\begin{align}
G^{(3)}_{ijk}(\boldsymbol{r}_{1},\boldsymbol{r}_{2},\boldsymbol{r}_{3}) & =\sum_{\lambda_{1}\lambda_{2}\lambda_{3}}\left\langle \tilde{\Psi} \right|\hat{\psi}_{i,\lambda_{1}}^{\dagger}(\boldsymbol{r}_{1})\hat{\psi}_{j,\lambda_{2}}^{\dagger}(\boldsymbol{r}_{2})\hat{\psi}_{k,\lambda_{3}}^{\dagger}(\boldsymbol{r}_{3})\hat{\psi}_{k,\lambda_{3}}(\boldsymbol{r}_{3})\hat{\psi}_{j,\lambda_{2}}(\boldsymbol{r}_{2})\hat{\psi}_{i,\lambda_{1}}(\boldsymbol{r}_{1})\left|\tilde{\Psi} \right\rangle ,\\
 & =\frac{1}{2}|\eta(\boldsymbol{r}_{1})|^{2}|\eta(\boldsymbol{r}_{2})|^{2}|\eta(\boldsymbol{r}_{3})|^{2}\left[\delta_{ij}(1-\delta_{jk})(1-\delta _{j,A}\delta_{k,B}-\delta _{j,B}\delta_{k,A}-\delta _{j,C}\delta_{k,D}-\delta _{j,D}\delta_{k,C})+|\epsilon_{ijk}|\right],
\end{align}
\begin{align}
G^{(4)}_{ijkl}(\boldsymbol{r}_{1},\boldsymbol{r}_{2},\boldsymbol{r}_{3},\boldsymbol{r}_{4}) & =\sum_{\lambda_{1}\lambda_{2}\lambda_{3}\lambda_{4}}\left\langle \tilde{\Psi} \right|\hat{\psi}_{i,\lambda_{1}}^{\dagger}(\boldsymbol{r}_{1})\hat{\psi}_{j,\lambda_{2}}^{\dagger}(\boldsymbol{r}_{2})\hat{\psi}_{k,\lambda_{3}}^{\dagger}(\boldsymbol{r}_{3})\hat{\psi}_{l,\lambda_{4}}^{\dagger}(\boldsymbol{r}_{4})\hat{\psi}_{l,\lambda_{4}}(\boldsymbol{r}_{4})\hat{\psi}_{k,\lambda_{3}}(\boldsymbol{r}_{3})\hat{\psi}_{j,\lambda_{2}}(\boldsymbol{r}_{2})\hat{\psi}_{i,\lambda_{1}}(\boldsymbol{r}_{1})\left|\tilde{\Psi} \right\rangle ,\\
 & =\frac{1}{4}|\eta(\boldsymbol{r}_{1})|^{2}|\eta(\boldsymbol{r}_{2})|^{2}|\eta(\boldsymbol{r}_{3})|^{2} |\eta(\boldsymbol{r}_{4})|^{2}  |\epsilon_{ijkl}|,
\end{align}
where $\epsilon_{ijk}$ and $\epsilon_{ijkl}$ are the Levi-Civita symbols. We observe that the information encoded by the imposed patterned phases cannot be extracted from these correlations. Similar to the two-photon case discussed in Section~\ref{ses:CoherenceContro2}, we can obtain the images corresponding to the four patterns from the following fourth-order correlation
\begin{align}
G^{(4)}_{ABCD}(\boldsymbol{r}_{1},\boldsymbol{r}_{2},\boldsymbol{r}_{3},\boldsymbol{r}_{4}) & =\left\langle \tilde{\Psi} \right|\hat{\psi}_{A,L}^{\dagger}(\boldsymbol{r}_{1})\hat{\psi}_{B,L}^{\dagger}(\boldsymbol{r}_{2})\hat{\psi}_{C,L}^{\dagger}(\boldsymbol{r}_{3})\hat{\psi}_{D,L}^{\dagger}(\boldsymbol{r}_{4})\hat{\psi}_{D,L}(\boldsymbol{r}_{4})\hat{\psi}_{C,L}(\boldsymbol{r}_{3})\hat{\psi}_{B,L}(\boldsymbol{r}_{2})\hat{\psi}_{A,L}(\boldsymbol{r}_{1})\left| \tilde{\Psi} \right\rangle ,\\
 & =\frac{1}{64}|\eta(\boldsymbol{r}_{1})|^{2}|\eta(\boldsymbol{r}_{2})|^{2}|\eta(\boldsymbol{r}_{3})|^{2} |\eta(\boldsymbol{r}_{4})|^{2}  \left\{ 1+\cos[\Phi_{A}(\boldsymbol{r}_{1})+\Phi_{B}(\boldsymbol{r}_{2})-\Phi_{C}(\boldsymbol{r}_{3}) -\Phi_{D}(\boldsymbol{r}_{4})]\right\}.
\end{align}
\end{widetext}
The $g^{(4)}$-function corresponding to the fourth-order correlation can be obtained as
\begin{equation}
  g^{(4)}_{ABCD}\!=\!\frac{1}{4}\left\{1\!+\!\cos[\Phi_{A}(\boldsymbol{r}_{1})\!+\!\Phi_{B}(\boldsymbol{r}_{2})\!-\!\Phi_{C}(\boldsymbol{r}_{3}) \!-\!\Phi_{D}(\boldsymbol{r}_{4})]\right\}. 
\end{equation}
These patterns encoded in the phases can only be observed in fourth-order correlations.

\section{Conclusion}\label{sec5}
We present a theoretical frame showcasing the accurate control of the spatial structure of the quantum coherence function through engineering the phases in their transverse plane. In addition to the helical structure of OAM photons~\cite{yang2022quantum}, we show that arbitrary patterns of phase profiles can be utilized to tail the coherence function for purposes. Furthermore, our method can be extended to higher-order coherence control. The exceptional controllability of quantum coherence in photons holds the potential for applications such as quantum correlated imaging and quantum sensing of angular rotations. Additionally, it serves as a driving force for advancing the development of the quantum version of structured illumination microscopy~\cite{heintzmann1999laterally,gustafsson2000surpassing}.

\section*{acknowledgments}
This work is supported by the National Natural Science Foundation of China (No. 12274215, No. 12175033, and No. 12275048); the Program for Innovative Talents and Entrepreneurs in Jiangsu; Key Research and Development Program of Guangdong Province (No. 2020B0303010001); and National Key R\&D Program of China (Grant No. 2021YFE0193500).

\bibliography{main}

\begin{thebibliography}{68}%
\makeatletter
\providecommand \@ifxundefined [1]{%
 \@ifx{#1\undefined}
}%
\providecommand \@ifnum [1]{%
 \ifnum #1\expandafter \@firstoftwo
 \else \expandafter \@secondoftwo
 \fi
}%
\providecommand \@ifx [1]{%
 \ifx #1\expandafter \@firstoftwo
 \else \expandafter \@secondoftwo
 \fi
}%
\providecommand \natexlab [1]{#1}%
\providecommand \enquote  [1]{``#1''}%
\providecommand \bibnamefont  [1]{#1}%
\providecommand \bibfnamefont [1]{#1}%
\providecommand \citenamefont [1]{#1}%
\providecommand \href@noop [0]{\@secondoftwo}%
\providecommand \href [0]{\begingroup \@sanitize@url \@href}%
\providecommand \@href[1]{\@@startlink{#1}\@@href}%
\providecommand \@@href[1]{\endgroup#1\@@endlink}%
\providecommand \@sanitize@url [0]{\catcode `\\12\catcode `\$12\catcode
  `\&12\catcode `\#12\catcode `\^12\catcode `\_12\catcode `\%12\relax}%
\providecommand \@@startlink[1]{}%
\providecommand \@@endlink[0]{}%
\providecommand \url  [0]{\begingroup\@sanitize@url \@url }%
\providecommand \@url [1]{\endgroup\@href {#1}{\urlprefix }}%
\providecommand \urlprefix  [0]{URL }%
\providecommand \Eprint [0]{\href }%
\providecommand \doibase [0]{https://doi.org/}%
\providecommand \selectlanguage [0]{\@gobble}%
\providecommand \bibinfo  [0]{\@secondoftwo}%
\providecommand \bibfield  [0]{\@secondoftwo}%
\providecommand \translation [1]{[#1]}%
\providecommand \BibitemOpen [0]{}%
\providecommand \bibitemStop [0]{}%
\providecommand \bibitemNoStop [0]{.\EOS\space}%
\providecommand \EOS [0]{\spacefactor3000\relax}%
\providecommand \BibitemShut  [1]{\csname bibitem#1\endcsname}%
\let\auto@bib@innerbib\@empty
\bibitem [{\citenamefont {Giovannetti}\ \emph {et~al.}(2004)\citenamefont
  {Giovannetti}, \citenamefont {Lloyd},\ and\ \citenamefont
  {Maccone}}]{giovannetti2004quantum}%
  \BibitemOpen
  \bibfield  {author} {\bibinfo {author} {\bibfnamefont {V.}~\bibnamefont
  {Giovannetti}}, \bibinfo {author} {\bibfnamefont {S.}~\bibnamefont {Lloyd}},\
  and\ \bibinfo {author} {\bibfnamefont {L.}~\bibnamefont {Maccone}},\
  }\bibfield  {title} {\bibinfo {title} {Quantum-enhanced measurements: beating
  the standard quantum limit},\ }\href
  {https://www.science.org/doi/abs/10.1126/science.1104149} {\bibfield
  {journal} {\bibinfo  {journal} {Science}\ }\textbf {\bibinfo {volume}
  {306}},\ \bibinfo {pages} {1330} (\bibinfo {year} {2004})}\BibitemShut
  {NoStop}%
\bibitem [{\citenamefont {Dowling}(2008)}]{dowling2008quantum}%
  \BibitemOpen
  \bibfield  {author} {\bibinfo {author} {\bibfnamefont {J.~P.}\ \bibnamefont
  {Dowling}},\ }\bibfield  {title} {\bibinfo {title} {Quantum optical
  metrology--the lowdown on high-n00n states},\ }\href
  {https://www.tandfonline.com/doi/abs/10.1080/00107510802091298} {\bibfield
  {journal} {\bibinfo  {journal} {Contemporary physics}\ }\textbf {\bibinfo
  {volume} {49}},\ \bibinfo {pages} {125} (\bibinfo {year} {2008})}\BibitemShut
  {NoStop}%
\bibitem [{\citenamefont {Lloyd}(2008)}]{lloyd2008enhanced}%
  \BibitemOpen
  \bibfield  {author} {\bibinfo {author} {\bibfnamefont {S.}~\bibnamefont
  {Lloyd}},\ }\bibfield  {title} {\bibinfo {title} {Enhanced sensitivity of
  photodetection via quantum illumination},\ }\href
  {https://www.science.org/doi/10.1126/science.1160627} {\bibfield  {journal}
  {\bibinfo  {journal} {Science}\ }\textbf {\bibinfo {volume} {321}},\ \bibinfo
  {pages} {1463} (\bibinfo {year} {2008})}\BibitemShut {NoStop}%
\bibitem [{\citenamefont {Walls}(1983)}]{walls1983squeezed}%
  \BibitemOpen
  \bibfield  {author} {\bibinfo {author} {\bibfnamefont {D.~F.}\ \bibnamefont
  {Walls}},\ }\bibfield  {title} {\bibinfo {title} {Squeezed states of light},\
  }\href {https://doi.org/10.1038/306141a0} {\bibfield  {journal} {\bibinfo
  {journal} {nature}\ }\textbf {\bibinfo {volume} {306}},\ \bibinfo {pages}
  {141} (\bibinfo {year} {1983})}\BibitemShut {NoStop}%
\bibitem [{\citenamefont {Wu}\ \emph {et~al.}(1986)\citenamefont {Wu},
  \citenamefont {Kimble}, \citenamefont {Hall},\ and\ \citenamefont
  {Wu}}]{wu1986generation}%
  \BibitemOpen
  \bibfield  {author} {\bibinfo {author} {\bibfnamefont {L.-A.}\ \bibnamefont
  {Wu}}, \bibinfo {author} {\bibfnamefont {H.~J.}\ \bibnamefont {Kimble}},
  \bibinfo {author} {\bibfnamefont {J.~L.}\ \bibnamefont {Hall}},\ and\
  \bibinfo {author} {\bibfnamefont {H.}~\bibnamefont {Wu}},\ }\bibfield
  {title} {\bibinfo {title} {Generation of squeezed states by parametric down
  conversion},\ }\href {https://doi.org/10.1103/PhysRevLett.57.2520} {\bibfield
   {journal} {\bibinfo  {journal} {Phys. Rev. Lett.}\ }\textbf {\bibinfo
  {volume} {57}},\ \bibinfo {pages} {2520} (\bibinfo {year}
  {1986})}\BibitemShut {NoStop}%
\bibitem [{\citenamefont {Slusher}\ \emph {et~al.}(1985)\citenamefont
  {Slusher}, \citenamefont {Hollberg}, \citenamefont {Yurke}, \citenamefont
  {Mertz},\ and\ \citenamefont {Valley}}]{slusher1985observation}%
  \BibitemOpen
  \bibfield  {author} {\bibinfo {author} {\bibfnamefont {R.~E.}\ \bibnamefont
  {Slusher}}, \bibinfo {author} {\bibfnamefont {L.~W.}\ \bibnamefont
  {Hollberg}}, \bibinfo {author} {\bibfnamefont {B.}~\bibnamefont {Yurke}},
  \bibinfo {author} {\bibfnamefont {J.~C.}\ \bibnamefont {Mertz}},\ and\
  \bibinfo {author} {\bibfnamefont {J.~F.}\ \bibnamefont {Valley}},\ }\bibfield
   {title} {\bibinfo {title} {Observation of squeezed states generated by
  four-wave mixing in an optical cavity},\ }\href
  {https://doi.org/10.1103/PhysRevLett.55.2409} {\bibfield  {journal} {\bibinfo
   {journal} {Phys. Rev. Lett.}\ }\textbf {\bibinfo {volume} {55}},\ \bibinfo
  {pages} {2409} (\bibinfo {year} {1985})}\BibitemShut {NoStop}%
\bibitem [{\citenamefont {Kok}\ \emph {et~al.}(2002)\citenamefont {Kok},
  \citenamefont {Lee},\ and\ \citenamefont {Dowling}}]{kok2002creation}%
  \BibitemOpen
  \bibfield  {author} {\bibinfo {author} {\bibfnamefont {P.}~\bibnamefont
  {Kok}}, \bibinfo {author} {\bibfnamefont {H.}~\bibnamefont {Lee}},\ and\
  \bibinfo {author} {\bibfnamefont {J.~P.}\ \bibnamefont {Dowling}},\
  }\bibfield  {title} {\bibinfo {title} {Creation of large-photon-number path
  entanglement conditioned on photodetection},\ }\href
  {https://doi.org/10.1103/PhysRevA.65.052104} {\bibfield  {journal} {\bibinfo
  {journal} {Phys. Rev. A}\ }\textbf {\bibinfo {volume} {65}},\ \bibinfo
  {pages} {052104} (\bibinfo {year} {2002})}\BibitemShut {NoStop}%
\bibitem [{\citenamefont {Afek}\ \emph {et~al.}(2010)\citenamefont {Afek},
  \citenamefont {Ambar},\ and\ \citenamefont {Silberberg}}]{afek2010high}%
  \BibitemOpen
  \bibfield  {author} {\bibinfo {author} {\bibfnamefont {I.}~\bibnamefont
  {Afek}}, \bibinfo {author} {\bibfnamefont {O.}~\bibnamefont {Ambar}},\ and\
  \bibinfo {author} {\bibfnamefont {Y.}~\bibnamefont {Silberberg}},\ }\bibfield
   {title} {\bibinfo {title} {High-noon states by mixing quantum and classical
  light},\ }\href {https://www.science.org/doi/10.1126/science.1188172}
  {\bibfield  {journal} {\bibinfo  {journal} {Science}\ }\textbf {\bibinfo
  {volume} {328}},\ \bibinfo {pages} {879} (\bibinfo {year}
  {2010})}\BibitemShut {NoStop}%
\bibitem [{\citenamefont {Kwiat}\ \emph {et~al.}(1999)\citenamefont {Kwiat},
  \citenamefont {Waks}, \citenamefont {White}, \citenamefont {Appelbaum},\ and\
  \citenamefont {Eberhard}}]{Kwiat1999ultrabright}%
  \BibitemOpen
  \bibfield  {author} {\bibinfo {author} {\bibfnamefont {P.~G.}\ \bibnamefont
  {Kwiat}}, \bibinfo {author} {\bibfnamefont {E.}~\bibnamefont {Waks}},
  \bibinfo {author} {\bibfnamefont {A.~G.}\ \bibnamefont {White}}, \bibinfo
  {author} {\bibfnamefont {I.}~\bibnamefont {Appelbaum}},\ and\ \bibinfo
  {author} {\bibfnamefont {P.~H.}\ \bibnamefont {Eberhard}},\ }\bibfield
  {title} {\bibinfo {title} {Ultrabright source of polarization-entangled
  photons},\ }\href {https://doi.org/10.1103/PhysRevA.60.R773} {\bibfield
  {journal} {\bibinfo  {journal} {Phys. Rev. A}\ }\textbf {\bibinfo {volume}
  {60}},\ \bibinfo {pages} {R773} (\bibinfo {year} {1999})}\BibitemShut
  {NoStop}%
\bibitem [{\citenamefont {Gisin}\ \emph {et~al.}(2002)\citenamefont {Gisin},
  \citenamefont {Ribordy}, \citenamefont {Tittel},\ and\ \citenamefont
  {Zbinden}}]{Gisin2002quantum}%
  \BibitemOpen
  \bibfield  {author} {\bibinfo {author} {\bibfnamefont {N.}~\bibnamefont
  {Gisin}}, \bibinfo {author} {\bibfnamefont {G.}~\bibnamefont {Ribordy}},
  \bibinfo {author} {\bibfnamefont {W.}~\bibnamefont {Tittel}},\ and\ \bibinfo
  {author} {\bibfnamefont {H.}~\bibnamefont {Zbinden}},\ }\bibfield  {title}
  {\bibinfo {title} {Quantum cryptography},\ }\href
  {https://doi.org/10.1103/RevModPhys.74.145} {\bibfield  {journal} {\bibinfo
  {journal} {Rev. Mod. Phys.}\ }\textbf {\bibinfo {volume} {74}},\ \bibinfo
  {pages} {145} (\bibinfo {year} {2002})}\BibitemShut {NoStop}%
\bibitem [{\citenamefont {Glauber}(1963)}]{Glauber1963}%
  \BibitemOpen
  \bibfield  {author} {\bibinfo {author} {\bibfnamefont {R.~J.}\ \bibnamefont
  {Glauber}},\ }\bibfield  {title} {\bibinfo {title} {The quantum theory of
  optical coherence},\ }\href {https://doi.org/10.1103/PhysRev.130.2529}
  {\bibfield  {journal} {\bibinfo  {journal} {Phys. Rev.}\ }\textbf {\bibinfo
  {volume} {130}},\ \bibinfo {pages} {2529} (\bibinfo {year}
  {1963})}\BibitemShut {NoStop}%
\bibitem [{\citenamefont {Brida}\ \emph {et~al.}(2010)\citenamefont {Brida},
  \citenamefont {Genovese},\ and\ \citenamefont
  {Berchera}}]{brida2010experimental}%
  \BibitemOpen
  \bibfield  {author} {\bibinfo {author} {\bibfnamefont {G.}~\bibnamefont
  {Brida}}, \bibinfo {author} {\bibfnamefont {M.}~\bibnamefont {Genovese}},\
  and\ \bibinfo {author} {\bibfnamefont {I.~R.}\ \bibnamefont {Berchera}},\
  }\bibfield  {title} {\bibinfo {title} {Experimental realization of
  sub-shot-noise quantum imaging},\ }\href
  {https://www.nature.com/articles/nphoton.2010.29} {\bibfield  {journal}
  {\bibinfo  {journal} {Nature Photonics}\ }\textbf {\bibinfo {volume} {4}},\
  \bibinfo {pages} {227} (\bibinfo {year} {2010})}\BibitemShut {NoStop}%
\bibitem [{\citenamefont {Ono}\ \emph {et~al.}(2013)\citenamefont {Ono},
  \citenamefont {Okamoto},\ and\ \citenamefont
  {Takeuchi}}]{ono2013entanglement}%
  \BibitemOpen
  \bibfield  {author} {\bibinfo {author} {\bibfnamefont {T.}~\bibnamefont
  {Ono}}, \bibinfo {author} {\bibfnamefont {R.}~\bibnamefont {Okamoto}},\ and\
  \bibinfo {author} {\bibfnamefont {S.}~\bibnamefont {Takeuchi}},\ }\bibfield
  {title} {\bibinfo {title} {An entanglement-enhanced microscope},\ }\href
  {https://doi.org/10.1038/ncomms3426} {\bibfield  {journal} {\bibinfo
  {journal} {Nature communications}\ }\textbf {\bibinfo {volume} {4}},\
  \bibinfo {pages} {2426} (\bibinfo {year} {2013})}\BibitemShut {NoStop}%
\bibitem [{\citenamefont {Gregory}\ \emph {et~al.}(2020)\citenamefont
  {Gregory}, \citenamefont {Moreau}, \citenamefont {Toninelli},\ and\
  \citenamefont {Padgett}}]{gregory2020imaging}%
  \BibitemOpen
  \bibfield  {author} {\bibinfo {author} {\bibfnamefont {T.}~\bibnamefont
  {Gregory}}, \bibinfo {author} {\bibfnamefont {P.-A.}\ \bibnamefont {Moreau}},
  \bibinfo {author} {\bibfnamefont {E.}~\bibnamefont {Toninelli}},\ and\
  \bibinfo {author} {\bibfnamefont {M.~J.}\ \bibnamefont {Padgett}},\
  }\bibfield  {title} {\bibinfo {title} {Imaging through noise with quantum
  illumination},\ }\href
  {https://www.science.org/doi/full/10.1126/sciadv.aay2652} {\bibfield
  {journal} {\bibinfo  {journal} {Science advances}\ }\textbf {\bibinfo
  {volume} {6}},\ \bibinfo {pages} {eaay2652} (\bibinfo {year}
  {2020})}\BibitemShut {NoStop}%
\bibitem [{\citenamefont {Moreau}\ \emph {et~al.}(2019)\citenamefont {Moreau},
  \citenamefont {Toninelli}, \citenamefont {Gregory},\ and\ \citenamefont
  {Padgett}}]{moreau2019imaging}%
  \BibitemOpen
  \bibfield  {author} {\bibinfo {author} {\bibfnamefont {P.-A.}\ \bibnamefont
  {Moreau}}, \bibinfo {author} {\bibfnamefont {E.}~\bibnamefont {Toninelli}},
  \bibinfo {author} {\bibfnamefont {T.}~\bibnamefont {Gregory}},\ and\ \bibinfo
  {author} {\bibfnamefont {M.~J.}\ \bibnamefont {Padgett}},\ }\bibfield
  {title} {\bibinfo {title} {Imaging with quantum states of light},\ }\href
  {https://doi.org/10.1038/s42254-019-0056-0} {\bibfield  {journal} {\bibinfo
  {journal} {Nature Reviews Physics}\ }\textbf {\bibinfo {volume} {1}},\
  \bibinfo {pages} {367} (\bibinfo {year} {2019})}\BibitemShut {NoStop}%
\bibitem [{\citenamefont {Wolfgramm}\ \emph {et~al.}(2013)\citenamefont
  {Wolfgramm}, \citenamefont {Vitelli}, \citenamefont {Beduini}, \citenamefont
  {Godbout},\ and\ \citenamefont {Mitchell}}]{wolfgramm2013entanglement}%
  \BibitemOpen
  \bibfield  {author} {\bibinfo {author} {\bibfnamefont {F.}~\bibnamefont
  {Wolfgramm}}, \bibinfo {author} {\bibfnamefont {C.}~\bibnamefont {Vitelli}},
  \bibinfo {author} {\bibfnamefont {F.~A.}\ \bibnamefont {Beduini}}, \bibinfo
  {author} {\bibfnamefont {N.}~\bibnamefont {Godbout}},\ and\ \bibinfo {author}
  {\bibfnamefont {M.~W.}\ \bibnamefont {Mitchell}},\ }\bibfield  {title}
  {\bibinfo {title} {Entanglement-enhanced probing of a delicate material
  system},\ }\href {https://doi.org/10.1038/nphoton.2012.300} {\bibfield
  {journal} {\bibinfo  {journal} {Nature Photonics}\ }\textbf {\bibinfo
  {volume} {7}},\ \bibinfo {pages} {28} (\bibinfo {year} {2013})}\BibitemShut
  {NoStop}%
\bibitem [{\citenamefont {Israel}\ \emph {et~al.}(2014)\citenamefont {Israel},
  \citenamefont {Rosen},\ and\ \citenamefont
  {Silberberg}}]{Israel2014supersensitive}%
  \BibitemOpen
  \bibfield  {author} {\bibinfo {author} {\bibfnamefont {Y.}~\bibnamefont
  {Israel}}, \bibinfo {author} {\bibfnamefont {S.}~\bibnamefont {Rosen}},\ and\
  \bibinfo {author} {\bibfnamefont {Y.}~\bibnamefont {Silberberg}},\ }\bibfield
   {title} {\bibinfo {title} {Supersensitive polarization microscopy using noon
  states of light},\ }\href {https://doi.org/10.1103/PhysRevLett.112.103604}
  {\bibfield  {journal} {\bibinfo  {journal} {Phys. Rev. Lett.}\ }\textbf
  {\bibinfo {volume} {112}},\ \bibinfo {pages} {103604} (\bibinfo {year}
  {2014})}\BibitemShut {NoStop}%
\bibitem [{\citenamefont {He}\ \emph {et~al.}(2023)\citenamefont {He},
  \citenamefont {Zhang}, \citenamefont {Tong}, \citenamefont {Li},\ and\
  \citenamefont {Wang}}]{he2023quantum}%
  \BibitemOpen
  \bibfield  {author} {\bibinfo {author} {\bibfnamefont {Z.}~\bibnamefont
  {He}}, \bibinfo {author} {\bibfnamefont {Y.}~\bibnamefont {Zhang}}, \bibinfo
  {author} {\bibfnamefont {X.}~\bibnamefont {Tong}}, \bibinfo {author}
  {\bibfnamefont {L.}~\bibnamefont {Li}},\ and\ \bibinfo {author}
  {\bibfnamefont {L.~V.}\ \bibnamefont {Wang}},\ }\bibfield  {title} {\bibinfo
  {title} {Quantum microscopy of cells at the heisenberg limit},\ }\href
  {https://doi.org/10.1038/s41467-023-38191-4} {\bibfield  {journal} {\bibinfo
  {journal} {Nature Communications}\ }\textbf {\bibinfo {volume} {14}},\
  \bibinfo {pages} {2441} (\bibinfo {year} {2023})}\BibitemShut {NoStop}%
\bibitem [{\citenamefont {Defienne}\ \emph {et~al.}(2021)\citenamefont
  {Defienne}, \citenamefont {Ndagano}, \citenamefont {Lyons},\ and\
  \citenamefont {Faccio}}]{Defienne2021polarization}%
  \BibitemOpen
  \bibfield  {author} {\bibinfo {author} {\bibfnamefont {H.}~\bibnamefont
  {Defienne}}, \bibinfo {author} {\bibfnamefont {B.}~\bibnamefont {Ndagano}},
  \bibinfo {author} {\bibfnamefont {A.}~\bibnamefont {Lyons}},\ and\ \bibinfo
  {author} {\bibfnamefont {D.}~\bibnamefont {Faccio}},\ }\bibfield  {title}
  {\bibinfo {title} {Polarization entanglement-enabled quantum holography},\
  }\href {https://doi.org/10.1038/s41567-020-01156-1} {\bibfield  {journal}
  {\bibinfo  {journal} {Nature Physics}\ }\textbf {\bibinfo {volume} {17}},\
  \bibinfo {pages} {591} (\bibinfo {year} {2021})}\BibitemShut {NoStop}%
\bibitem [{\citenamefont {Camphausen}\ \emph {et~al.}(2021)\citenamefont
  {Camphausen}, \citenamefont {Álvaro Cuevas}, \citenamefont {Duempelmann},
  \citenamefont {Terborg}, \citenamefont {Wajs}, \citenamefont {Tisa},
  \citenamefont {Ruggeri}, \citenamefont {Cusini}, \citenamefont
  {Steinlechner},\ and\ \citenamefont {Pruneri}}]{Camphausen2021quantum}%
  \BibitemOpen
  \bibfield  {author} {\bibinfo {author} {\bibfnamefont {R.}~\bibnamefont
  {Camphausen}}, \bibinfo {author} {\bibnamefont {Álvaro Cuevas}}, \bibinfo
  {author} {\bibfnamefont {L.}~\bibnamefont {Duempelmann}}, \bibinfo {author}
  {\bibfnamefont {R.~A.}\ \bibnamefont {Terborg}}, \bibinfo {author}
  {\bibfnamefont {E.}~\bibnamefont {Wajs}}, \bibinfo {author} {\bibfnamefont
  {S.}~\bibnamefont {Tisa}}, \bibinfo {author} {\bibfnamefont {A.}~\bibnamefont
  {Ruggeri}}, \bibinfo {author} {\bibfnamefont {I.}~\bibnamefont {Cusini}},
  \bibinfo {author} {\bibfnamefont {F.}~\bibnamefont {Steinlechner}},\ and\
  \bibinfo {author} {\bibfnamefont {V.}~\bibnamefont {Pruneri}},\ }\bibfield
  {title} {\bibinfo {title} {A quantum-enhanced wide-field phase imager},\
  }\href {https://doi.org/10.1126/sciadv.abj2155} {\bibfield  {journal}
  {\bibinfo  {journal} {Science Advances}\ }\textbf {\bibinfo {volume} {7}},\
  \bibinfo {pages} {eabj2155} (\bibinfo {year} {2021})}\BibitemShut {NoStop}%
\bibitem [{\citenamefont {Black}\ \emph {et~al.}(2023)\citenamefont {Black},
  \citenamefont {Nguyen}, \citenamefont {Braverman}, \citenamefont {Crampton},
  \citenamefont {Evans},\ and\ \citenamefont {Boyd}}]{Black23quantum}%
  \BibitemOpen
  \bibfield  {author} {\bibinfo {author} {\bibfnamefont {A.~N.}\ \bibnamefont
  {Black}}, \bibinfo {author} {\bibfnamefont {L.~D.}\ \bibnamefont {Nguyen}},
  \bibinfo {author} {\bibfnamefont {B.}~\bibnamefont {Braverman}}, \bibinfo
  {author} {\bibfnamefont {K.~T.}\ \bibnamefont {Crampton}}, \bibinfo {author}
  {\bibfnamefont {J.~E.}\ \bibnamefont {Evans}},\ and\ \bibinfo {author}
  {\bibfnamefont {R.~W.}\ \bibnamefont {Boyd}},\ }\bibfield  {title} {\bibinfo
  {title} {Quantum-enhanced phase imaging without coincidence counting},\
  }\href {https://doi.org/10.1364/OPTICA.482926} {\bibfield  {journal}
  {\bibinfo  {journal} {Optica}\ }\textbf {\bibinfo {volume} {10}},\ \bibinfo
  {pages} {952} (\bibinfo {year} {2023})}\BibitemShut {NoStop}%
\bibitem [{\citenamefont {Chrapkiewicz}\ \emph {et~al.}(2016)\citenamefont
  {Chrapkiewicz}, \citenamefont {Jachura}, \citenamefont {Banaszek},\ and\
  \citenamefont {Wasilewski}}]{chrapkiewicz2016hologram}%
  \BibitemOpen
  \bibfield  {author} {\bibinfo {author} {\bibfnamefont {R.}~\bibnamefont
  {Chrapkiewicz}}, \bibinfo {author} {\bibfnamefont {M.}~\bibnamefont
  {Jachura}}, \bibinfo {author} {\bibfnamefont {K.}~\bibnamefont {Banaszek}},\
  and\ \bibinfo {author} {\bibfnamefont {W.}~\bibnamefont {Wasilewski}},\
  }\bibfield  {title} {\bibinfo {title} {Hologram of a single photon},\ }\href
  {https://www.nature.com/articles/nphoton.2016.129} {\bibfield  {journal}
  {\bibinfo  {journal} {Nature Photonics}\ }\textbf {\bibinfo {volume} {10}},\
  \bibinfo {pages} {576} (\bibinfo {year} {2016})}\BibitemShut {NoStop}%
\bibitem [{\citenamefont {Ndagano}\ \emph {et~al.}(2022)\citenamefont
  {Ndagano}, \citenamefont {Defienne}, \citenamefont {Branford}, \citenamefont
  {Shah}, \citenamefont {Lyons}, \citenamefont {Westerberg}, \citenamefont
  {Gauger},\ and\ \citenamefont {Faccio}}]{Ndagano2022microscopy}%
  \BibitemOpen
  \bibfield  {author} {\bibinfo {author} {\bibfnamefont {B.}~\bibnamefont
  {Ndagano}}, \bibinfo {author} {\bibfnamefont {H.}~\bibnamefont {Defienne}},
  \bibinfo {author} {\bibfnamefont {D.}~\bibnamefont {Branford}}, \bibinfo
  {author} {\bibfnamefont {Y.~D.}\ \bibnamefont {Shah}}, \bibinfo {author}
  {\bibfnamefont {A.}~\bibnamefont {Lyons}}, \bibinfo {author} {\bibfnamefont
  {N.}~\bibnamefont {Westerberg}}, \bibinfo {author} {\bibfnamefont {E.~M.}\
  \bibnamefont {Gauger}},\ and\ \bibinfo {author} {\bibfnamefont
  {D.}~\bibnamefont {Faccio}},\ }\bibfield  {title} {\bibinfo {title} {Quantum
  microscopy based on hong--ou--mandel interference},\ }\href
  {https://doi.org/10.1038/s41566-022-00980-6} {\bibfield  {journal} {\bibinfo
  {journal} {Nature Photonics}\ }\textbf {\bibinfo {volume} {16}},\ \bibinfo
  {pages} {384} (\bibinfo {year} {2022})}\BibitemShut {NoStop}%
\bibitem [{\citenamefont {Zia}\ \emph {et~al.}(2023)\citenamefont {Zia},
  \citenamefont {Dehghan}, \citenamefont {D’Errico}, \citenamefont
  {Sciarrino},\ and\ \citenamefont {Karimi}}]{zia2023interferometric}%
  \BibitemOpen
  \bibfield  {author} {\bibinfo {author} {\bibfnamefont {D.}~\bibnamefont
  {Zia}}, \bibinfo {author} {\bibfnamefont {N.}~\bibnamefont {Dehghan}},
  \bibinfo {author} {\bibfnamefont {A.}~\bibnamefont {D’Errico}}, \bibinfo
  {author} {\bibfnamefont {F.}~\bibnamefont {Sciarrino}},\ and\ \bibinfo
  {author} {\bibfnamefont {E.}~\bibnamefont {Karimi}},\ }\bibfield  {title}
  {\bibinfo {title} {Interferometric imaging of amplitude and phase of spatial
  biphoton states},\ }\href {https://doi.org/10.1038/s41566-023-01272-3}
  {\bibfield  {journal} {\bibinfo  {journal} {Nature Photonics}\ ,\ \bibinfo
  {pages} {1}} (\bibinfo {year} {2023})}\BibitemShut {NoStop}%
\bibitem [{\citenamefont {Walborn}\ \emph {et~al.}(2010)\citenamefont
  {Walborn}, \citenamefont {Monken}, \citenamefont {P{\'a}dua},\ and\
  \citenamefont {Ribeiro}}]{walborn2010spatial}%
  \BibitemOpen
  \bibfield  {author} {\bibinfo {author} {\bibfnamefont {S.~P.}\ \bibnamefont
  {Walborn}}, \bibinfo {author} {\bibfnamefont {C.}~\bibnamefont {Monken}},
  \bibinfo {author} {\bibfnamefont {S.}~\bibnamefont {P{\'a}dua}},\ and\
  \bibinfo {author} {\bibfnamefont {P.~S.}\ \bibnamefont {Ribeiro}},\
  }\bibfield  {title} {\bibinfo {title} {Spatial correlations in parametric
  down-conversion},\ }\href {https://doi.org/10.1016/j.physrep.2010.06.003}
  {\bibfield  {journal} {\bibinfo  {journal} {Physics Reports}\ }\textbf
  {\bibinfo {volume} {495}},\ \bibinfo {pages} {87} (\bibinfo {year}
  {2010})}\BibitemShut {NoStop}%
\bibitem [{\citenamefont {Law}\ and\ \citenamefont
  {Eberly}(2004)}]{Law2004analysis}%
  \BibitemOpen
  \bibfield  {author} {\bibinfo {author} {\bibfnamefont {C.~K.}\ \bibnamefont
  {Law}}\ and\ \bibinfo {author} {\bibfnamefont {J.~H.}\ \bibnamefont
  {Eberly}},\ }\bibfield  {title} {\bibinfo {title} {Analysis and
  interpretation of high transverse entanglement in optical parametric down
  conversion},\ }\href {https://doi.org/10.1103/PhysRevLett.92.127903}
  {\bibfield  {journal} {\bibinfo  {journal} {Phys. Rev. Lett.}\ }\textbf
  {\bibinfo {volume} {92}},\ \bibinfo {pages} {127903} (\bibinfo {year}
  {2004})}\BibitemShut {NoStop}%
\bibitem [{\citenamefont {Yu}\ \emph {et~al.}(2011)\citenamefont {Yu},
  \citenamefont {Genevet}, \citenamefont {Kats}, \citenamefont {Aieta},
  \citenamefont {Tetienne}, \citenamefont {Capasso},\ and\ \citenamefont
  {Gaburro}}]{Yu2011Light}%
  \BibitemOpen
  \bibfield  {author} {\bibinfo {author} {\bibfnamefont {N.}~\bibnamefont
  {Yu}}, \bibinfo {author} {\bibfnamefont {P.}~\bibnamefont {Genevet}},
  \bibinfo {author} {\bibfnamefont {M.~A.}\ \bibnamefont {Kats}}, \bibinfo
  {author} {\bibfnamefont {F.}~\bibnamefont {Aieta}}, \bibinfo {author}
  {\bibfnamefont {J.-P.}\ \bibnamefont {Tetienne}}, \bibinfo {author}
  {\bibfnamefont {F.}~\bibnamefont {Capasso}},\ and\ \bibinfo {author}
  {\bibfnamefont {Z.}~\bibnamefont {Gaburro}},\ }\bibfield  {title} {\bibinfo
  {title} {Light propagation with phase discontinuities: Generalized laws of
  reflection and refraction},\ }\href {https://doi.org/10.1126/science.1210713}
  {\bibfield  {journal} {\bibinfo  {journal} {Science}\ }\textbf {\bibinfo
  {volume} {334}},\ \bibinfo {pages} {333} (\bibinfo {year}
  {2011})}\BibitemShut {NoStop}%
\bibitem [{\citenamefont {Devlin}\ \emph {et~al.}(2017)\citenamefont {Devlin},
  \citenamefont {Ambrosio}, \citenamefont {Rubin}, \citenamefont {Mueller},\
  and\ \citenamefont {Capasso}}]{devlin2017Arbitrary}%
  \BibitemOpen
  \bibfield  {author} {\bibinfo {author} {\bibfnamefont {R.~C.}\ \bibnamefont
  {Devlin}}, \bibinfo {author} {\bibfnamefont {A.}~\bibnamefont {Ambrosio}},
  \bibinfo {author} {\bibfnamefont {N.~A.}\ \bibnamefont {Rubin}}, \bibinfo
  {author} {\bibfnamefont {J.~P.~B.}\ \bibnamefont {Mueller}},\ and\ \bibinfo
  {author} {\bibfnamefont {F.}~\bibnamefont {Capasso}},\ }\bibfield  {title}
  {\bibinfo {title} {Arbitrary spin-to–orbital angular momentum conversion of
  light},\ }\href {https://doi.org/10.1126/science.aao5392} {\bibfield
  {journal} {\bibinfo  {journal} {Science}\ }\textbf {\bibinfo {volume}
  {358}},\ \bibinfo {pages} {896} (\bibinfo {year} {2017})}\BibitemShut
  {NoStop}%
\bibitem [{\citenamefont {Shen}\ \emph {et~al.}(2019)\citenamefont {Shen},
  \citenamefont {Wang}, \citenamefont {Xie}, \citenamefont {Min}, \citenamefont
  {Fu}, \citenamefont {Liu}, \citenamefont {Gong},\ and\ \citenamefont
  {Yuan}}]{Shen2019}%
  \BibitemOpen
  \bibfield  {author} {\bibinfo {author} {\bibfnamefont {Y.}~\bibnamefont
  {Shen}}, \bibinfo {author} {\bibfnamefont {X.}~\bibnamefont {Wang}}, \bibinfo
  {author} {\bibfnamefont {Z.}~\bibnamefont {Xie}}, \bibinfo {author}
  {\bibfnamefont {C.}~\bibnamefont {Min}}, \bibinfo {author} {\bibfnamefont
  {X.}~\bibnamefont {Fu}}, \bibinfo {author} {\bibfnamefont {Q.}~\bibnamefont
  {Liu}}, \bibinfo {author} {\bibfnamefont {M.}~\bibnamefont {Gong}},\ and\
  \bibinfo {author} {\bibfnamefont {X.}~\bibnamefont {Yuan}},\ }\bibfield
  {title} {\bibinfo {title} {Optical vortices 30 years on: {OAM} manipulation
  from topological charge to multiple singularities},\ }\href
  {https://doi.org/10.1038/s41377-019-0194-2} {\bibfield  {journal} {\bibinfo
  {journal} {Light: Science {\&} Applications}\ }\textbf {\bibinfo {volume}
  {8}},\ \bibinfo {pages} {90} (\bibinfo {year} {2019})}\BibitemShut {NoStop}%
\bibitem [{\citenamefont {Forbes}(2020)}]{forbes2020structured}%
  \BibitemOpen
  \bibfield  {author} {\bibinfo {author} {\bibfnamefont {A.}~\bibnamefont
  {Forbes}},\ }\bibfield  {title} {\bibinfo {title} {Structured light: tailored
  for purpose},\ }\href
  {https://www.optica-opn.org/home/articles/volume_31/june_2020/features/structured_light_tailored_for_purpose/?src=hpeditor}
  {\bibfield  {journal} {\bibinfo  {journal} {Opt. Photon. News}\ }\textbf
  {\bibinfo {volume} {31}},\ \bibinfo {pages} {24} (\bibinfo {year}
  {2020})}\BibitemShut {NoStop}%
\bibitem [{\citenamefont {Nagali}\ \emph {et~al.}(2009)\citenamefont {Nagali},
  \citenamefont {Sciarrino}, \citenamefont {De~Martini}, \citenamefont
  {Marrucci}, \citenamefont {Piccirillo}, \citenamefont {Karimi},\ and\
  \citenamefont {Santamato}}]{Nagali2009quantum}%
  \BibitemOpen
  \bibfield  {author} {\bibinfo {author} {\bibfnamefont {E.}~\bibnamefont
  {Nagali}}, \bibinfo {author} {\bibfnamefont {F.}~\bibnamefont {Sciarrino}},
  \bibinfo {author} {\bibfnamefont {F.}~\bibnamefont {De~Martini}}, \bibinfo
  {author} {\bibfnamefont {L.}~\bibnamefont {Marrucci}}, \bibinfo {author}
  {\bibfnamefont {B.}~\bibnamefont {Piccirillo}}, \bibinfo {author}
  {\bibfnamefont {E.}~\bibnamefont {Karimi}},\ and\ \bibinfo {author}
  {\bibfnamefont {E.}~\bibnamefont {Santamato}},\ }\bibfield  {title} {\bibinfo
  {title} {Quantum information transfer from spin to orbital angular momentum
  of photons},\ }\href {https://doi.org/10.1103/PhysRevLett.103.013601}
  {\bibfield  {journal} {\bibinfo  {journal} {Phys. Rev. Lett.}\ }\textbf
  {\bibinfo {volume} {103}},\ \bibinfo {pages} {013601} (\bibinfo {year}
  {2009})}\BibitemShut {NoStop}%
\bibitem [{\citenamefont {Stav}\ \emph {et~al.}(2018)\citenamefont {Stav},
  \citenamefont {Faerman}, \citenamefont {Maguid}, \citenamefont {Oren},
  \citenamefont {Kleiner}, \citenamefont {Hasman},\ and\ \citenamefont
  {Segev}}]{stav2018quantum}%
  \BibitemOpen
  \bibfield  {author} {\bibinfo {author} {\bibfnamefont {T.}~\bibnamefont
  {Stav}}, \bibinfo {author} {\bibfnamefont {A.}~\bibnamefont {Faerman}},
  \bibinfo {author} {\bibfnamefont {E.}~\bibnamefont {Maguid}}, \bibinfo
  {author} {\bibfnamefont {D.}~\bibnamefont {Oren}}, \bibinfo {author}
  {\bibfnamefont {V.}~\bibnamefont {Kleiner}}, \bibinfo {author} {\bibfnamefont
  {E.}~\bibnamefont {Hasman}},\ and\ \bibinfo {author} {\bibfnamefont
  {M.}~\bibnamefont {Segev}},\ }\bibfield  {title} {\bibinfo {title} {Quantum
  entanglement of the spin and orbital angular momentum of photons using
  metamaterials},\ }\href
  {https://science.sciencemag.org/content/361/6407/1101} {\bibfield  {journal}
  {\bibinfo  {journal} {Science}\ }\textbf {\bibinfo {volume} {361}},\ \bibinfo
  {pages} {1101} (\bibinfo {year} {2018})}\BibitemShut {NoStop}%
\bibitem [{\citenamefont {Morris}\ \emph {et~al.}(2015)\citenamefont {Morris},
  \citenamefont {Aspden}, \citenamefont {Bell}, \citenamefont {Boyd},\ and\
  \citenamefont {Padgett}}]{morris2015imaging}%
  \BibitemOpen
  \bibfield  {author} {\bibinfo {author} {\bibfnamefont {P.~A.}\ \bibnamefont
  {Morris}}, \bibinfo {author} {\bibfnamefont {R.~S.}\ \bibnamefont {Aspden}},
  \bibinfo {author} {\bibfnamefont {J.~E.}\ \bibnamefont {Bell}}, \bibinfo
  {author} {\bibfnamefont {R.~W.}\ \bibnamefont {Boyd}},\ and\ \bibinfo
  {author} {\bibfnamefont {M.~J.}\ \bibnamefont {Padgett}},\ }\bibfield
  {title} {\bibinfo {title} {Imaging with a small number of photons},\ }\href
  {https://www.nature.com/articles/ncomms6913} {\bibfield  {journal} {\bibinfo
  {journal} {Nature communications}\ }\textbf {\bibinfo {volume} {6}},\
  \bibinfo {pages} {1} (\bibinfo {year} {2015})}\BibitemShut {NoStop}%
\bibitem [{\citenamefont {Lemos}\ \emph {et~al.}(2014)\citenamefont {Lemos},
  \citenamefont {Borish}, \citenamefont {Cole}, \citenamefont {Ramelow},
  \citenamefont {Lapkiewicz},\ and\ \citenamefont
  {Zeilinger}}]{lemos2014quantum}%
  \BibitemOpen
  \bibfield  {author} {\bibinfo {author} {\bibfnamefont {G.~B.}\ \bibnamefont
  {Lemos}}, \bibinfo {author} {\bibfnamefont {V.}~\bibnamefont {Borish}},
  \bibinfo {author} {\bibfnamefont {G.~D.}\ \bibnamefont {Cole}}, \bibinfo
  {author} {\bibfnamefont {S.}~\bibnamefont {Ramelow}}, \bibinfo {author}
  {\bibfnamefont {R.}~\bibnamefont {Lapkiewicz}},\ and\ \bibinfo {author}
  {\bibfnamefont {A.}~\bibnamefont {Zeilinger}},\ }\bibfield  {title} {\bibinfo
  {title} {Quantum imaging with undetected photons},\ }\href
  {https://www.nature.com/articles/nature13586} {\bibfield  {journal} {\bibinfo
   {journal} {Nature}\ }\textbf {\bibinfo {volume} {512}},\ \bibinfo {pages}
  {409} (\bibinfo {year} {2014})}\BibitemShut {NoStop}%
\bibitem [{\citenamefont {Maga{\~n}a-Loaiza}\ and\ \citenamefont
  {Boyd}(2019)}]{magana2019quantum}%
  \BibitemOpen
  \bibfield  {author} {\bibinfo {author} {\bibfnamefont {O.~S.}\ \bibnamefont
  {Maga{\~n}a-Loaiza}}\ and\ \bibinfo {author} {\bibfnamefont {R.~W.}\
  \bibnamefont {Boyd}},\ }\bibfield  {title} {\bibinfo {title} {Quantum imaging
  and information},\ }\href
  {https://iopscience.iop.org/article/10.1088/1361-6633/ab5005/pdf} {\bibfield
  {journal} {\bibinfo  {journal} {Reports on Progress in Physics}\ }\textbf
  {\bibinfo {volume} {82}},\ \bibinfo {pages} {124401} (\bibinfo {year}
  {2019})}\BibitemShut {NoStop}%
\bibitem [{\citenamefont {Lavery}\ \emph {et~al.}(2013)\citenamefont {Lavery},
  \citenamefont {Speirits}, \citenamefont {Barnett},\ and\ \citenamefont
  {Padgett}}]{Lavery2013detection}%
  \BibitemOpen
  \bibfield  {author} {\bibinfo {author} {\bibfnamefont {M.~P.~J.}\
  \bibnamefont {Lavery}}, \bibinfo {author} {\bibfnamefont {F.~C.}\
  \bibnamefont {Speirits}}, \bibinfo {author} {\bibfnamefont {S.~M.}\
  \bibnamefont {Barnett}},\ and\ \bibinfo {author} {\bibfnamefont {M.~J.}\
  \bibnamefont {Padgett}},\ }\bibfield  {title} {\bibinfo {title} {Detection of
  a spinning object using light’s orbital angular momentum},\ }\href
  {https://doi.org/10.1126/science.1239936} {\bibfield  {journal} {\bibinfo
  {journal} {Science}\ }\textbf {\bibinfo {volume} {341}},\ \bibinfo {pages}
  {537} (\bibinfo {year} {2013})}\BibitemShut {NoStop}%
\bibitem [{\citenamefont {Korech}\ \emph {et~al.}(2013)\citenamefont {Korech},
  \citenamefont {Steinitz}, \citenamefont {Gordon}, \citenamefont {Averbukh},\
  and\ \citenamefont {Prior}}]{Korech2013}%
  \BibitemOpen
  \bibfield  {author} {\bibinfo {author} {\bibfnamefont {O.}~\bibnamefont
  {Korech}}, \bibinfo {author} {\bibfnamefont {U.}~\bibnamefont {Steinitz}},
  \bibinfo {author} {\bibfnamefont {R.~J.}\ \bibnamefont {Gordon}}, \bibinfo
  {author} {\bibfnamefont {I.~S.}\ \bibnamefont {Averbukh}},\ and\ \bibinfo
  {author} {\bibfnamefont {Y.}~\bibnamefont {Prior}},\ }\bibfield  {title}
  {\bibinfo {title} {Observing molecular spinning via the rotational doppler
  effect},\ }\href {https://doi.org/10.1038/nphoton.2013.189} {\bibfield
  {journal} {\bibinfo  {journal} {Nature Photonics}\ }\textbf {\bibinfo
  {volume} {7}},\ \bibinfo {pages} {711} (\bibinfo {year} {2013})}\BibitemShut
  {NoStop}%
\bibitem [{\citenamefont {Zhang}\ \emph {et~al.}(2019)\citenamefont {Zhang},
  \citenamefont {Zhang}, \citenamefont {Qiu},\ and\ \citenamefont
  {Chen}}]{Chen2019Quantum}%
  \BibitemOpen
  \bibfield  {author} {\bibinfo {author} {\bibfnamefont {W.}~\bibnamefont
  {Zhang}}, \bibinfo {author} {\bibfnamefont {D.}~\bibnamefont {Zhang}},
  \bibinfo {author} {\bibfnamefont {X.}~\bibnamefont {Qiu}},\ and\ \bibinfo
  {author} {\bibfnamefont {L.}~\bibnamefont {Chen}},\ }\bibfield  {title}
  {\bibinfo {title} {Quantum remote sensing of the angular rotation of
  structured objects},\ }\href {https://doi.org/10.1103/PhysRevA.100.043832}
  {\bibfield  {journal} {\bibinfo  {journal} {Phys. Rev. A}\ }\textbf {\bibinfo
  {volume} {100}},\ \bibinfo {pages} {043832} (\bibinfo {year}
  {2019})}\BibitemShut {NoStop}%
\bibitem [{\citenamefont {Qiu}\ \emph {et~al.}(2022)\citenamefont {Qiu},
  \citenamefont {Ding}, \citenamefont {Liu}, \citenamefont {Liu}, \citenamefont
  {Wu},\ and\ \citenamefont {Ren}}]{qiu2022fragmental}%
  \BibitemOpen
  \bibfield  {author} {\bibinfo {author} {\bibfnamefont {S.}~\bibnamefont
  {Qiu}}, \bibinfo {author} {\bibfnamefont {Y.}~\bibnamefont {Ding}}, \bibinfo
  {author} {\bibfnamefont {T.}~\bibnamefont {Liu}}, \bibinfo {author}
  {\bibfnamefont {Z.}~\bibnamefont {Liu}}, \bibinfo {author} {\bibfnamefont
  {H.}~\bibnamefont {Wu}},\ and\ \bibinfo {author} {\bibfnamefont
  {Y.}~\bibnamefont {Ren}},\ }\bibfield  {title} {\bibinfo {title} {Fragmental
  optical vortex for the detection of rotating object based on the rotational
  doppler effect},\ }\href {https://doi.org/10.1364/OE.476870} {\bibfield
  {journal} {\bibinfo  {journal} {Optics Express}\ }\textbf {\bibinfo {volume}
  {30}},\ \bibinfo {pages} {47350} (\bibinfo {year} {2022})}\BibitemShut
  {NoStop}%
\bibitem [{\citenamefont {Yang}\ and\ \citenamefont
  {Xu}(2022)}]{yang2022quantum}%
  \BibitemOpen
  \bibfield  {author} {\bibinfo {author} {\bibfnamefont {L.-P.}\ \bibnamefont
  {Yang}}\ and\ \bibinfo {author} {\bibfnamefont {D.}~\bibnamefont {Xu}},\
  }\bibfield  {title} {\bibinfo {title} {Quantum theory of photonic vortices
  and quantum statistics of twisted photons},\ }\href
  {https://doi.org/10.1103/PhysRevA.105.023723} {\bibfield  {journal} {\bibinfo
   {journal} {Phys. Rev. A}\ }\textbf {\bibinfo {volume} {105}},\ \bibinfo
  {pages} {023723} (\bibinfo {year} {2022})}\BibitemShut {NoStop}%
\bibitem [{\citenamefont {Gao}\ \emph {et~al.}(2023)\citenamefont {Gao},
  \citenamefont {Zhang}, \citenamefont {D'Errico}, \citenamefont {Sit},
  \citenamefont {Heshami},\ and\ \citenamefont {Karimi}}]{gao2023full}%
  \BibitemOpen
  \bibfield  {author} {\bibinfo {author} {\bibfnamefont {X.}~\bibnamefont
  {Gao}}, \bibinfo {author} {\bibfnamefont {Y.}~\bibnamefont {Zhang}}, \bibinfo
  {author} {\bibfnamefont {A.}~\bibnamefont {D'Errico}}, \bibinfo {author}
  {\bibfnamefont {A.}~\bibnamefont {Sit}}, \bibinfo {author} {\bibfnamefont
  {K.}~\bibnamefont {Heshami}},\ and\ \bibinfo {author} {\bibfnamefont
  {E.}~\bibnamefont {Karimi}},\ }\bibfield  {title} {\bibinfo {title} {Full
  spatial characterization of entangled structured photons},\ }\href
  {https://arxiv.org/abs/2304.14280} {\bibfield  {journal} {\bibinfo  {journal}
  {arXiv preprint arXiv:2304.14280}\ } (\bibinfo {year} {2023})}\BibitemShut
  {NoStop}%
\bibitem [{\citenamefont {Huang}\ \emph {et~al.}(2023)\citenamefont {Huang},
  \citenamefont {Gao}, \citenamefont {Ren}, \citenamefont {Cheng},
  \citenamefont {Zhu}, \citenamefont {Xue}, \citenamefont {Lou}, \citenamefont
  {Liu}, \citenamefont {Chen}, \citenamefont {Zhu} \emph
  {et~al.}}]{huang2023manipulating}%
  \BibitemOpen
  \bibfield  {author} {\bibinfo {author} {\bibfnamefont {S.-Y.}\ \bibnamefont
  {Huang}}, \bibinfo {author} {\bibfnamefont {J.}~\bibnamefont {Gao}}, \bibinfo
  {author} {\bibfnamefont {Z.-C.}\ \bibnamefont {Ren}}, \bibinfo {author}
  {\bibfnamefont {Z.-M.}\ \bibnamefont {Cheng}}, \bibinfo {author}
  {\bibfnamefont {W.-Z.}\ \bibnamefont {Zhu}}, \bibinfo {author} {\bibfnamefont
  {S.-T.}\ \bibnamefont {Xue}}, \bibinfo {author} {\bibfnamefont {Y.-C.}\
  \bibnamefont {Lou}}, \bibinfo {author} {\bibfnamefont {Z.-F.}\ \bibnamefont
  {Liu}}, \bibinfo {author} {\bibfnamefont {C.}~\bibnamefont {Chen}}, \bibinfo
  {author} {\bibfnamefont {F.}~\bibnamefont {Zhu}}, \emph {et~al.},\ }\bibfield
   {title} {\bibinfo {title} {Manipulating spatial structure of high-order
  quantum coherence with entangled photons},\ }\href
  {https://arxiv.org/abs/2306.00772} {\bibfield  {journal} {\bibinfo  {journal}
  {arXiv preprint arXiv:2306.00772}\ } (\bibinfo {year} {2023})}\BibitemShut
  {NoStop}%
\bibitem [{\citenamefont {Gilaberte~Basset}\ \emph {et~al.}(2019)\citenamefont
  {Gilaberte~Basset}, \citenamefont {Setzpfandt}, \citenamefont {Steinlechner},
  \citenamefont {Beckert}, \citenamefont {Pertsch},\ and\ \citenamefont
  {Gr{\"a}fe}}]{gilaberte2019perspectives}%
  \BibitemOpen
  \bibfield  {author} {\bibinfo {author} {\bibfnamefont {M.}~\bibnamefont
  {Gilaberte~Basset}}, \bibinfo {author} {\bibfnamefont {F.}~\bibnamefont
  {Setzpfandt}}, \bibinfo {author} {\bibfnamefont {F.}~\bibnamefont
  {Steinlechner}}, \bibinfo {author} {\bibfnamefont {E.}~\bibnamefont
  {Beckert}}, \bibinfo {author} {\bibfnamefont {T.}~\bibnamefont {Pertsch}},\
  and\ \bibinfo {author} {\bibfnamefont {M.}~\bibnamefont {Gr{\"a}fe}},\
  }\bibfield  {title} {\bibinfo {title} {Perspectives for applications of
  quantum imaging},\ }\href {https://doi.org/10.1002/lpor.201900097} {\bibfield
   {journal} {\bibinfo  {journal} {Laser \& Photonics Reviews}\ }\textbf
  {\bibinfo {volume} {13}},\ \bibinfo {pages} {1900097} (\bibinfo {year}
  {2019})}\BibitemShut {NoStop}%
\bibitem [{\citenamefont {Taylor}\ \emph {et~al.}(2013)\citenamefont {Taylor},
  \citenamefont {Janousek}, \citenamefont {Daria}, \citenamefont {Knittel},
  \citenamefont {Hage}, \citenamefont {Bachor},\ and\ \citenamefont
  {Bowen}}]{taylor2013biological}%
  \BibitemOpen
  \bibfield  {author} {\bibinfo {author} {\bibfnamefont {M.~A.}\ \bibnamefont
  {Taylor}}, \bibinfo {author} {\bibfnamefont {J.}~\bibnamefont {Janousek}},
  \bibinfo {author} {\bibfnamefont {V.}~\bibnamefont {Daria}}, \bibinfo
  {author} {\bibfnamefont {J.}~\bibnamefont {Knittel}}, \bibinfo {author}
  {\bibfnamefont {B.}~\bibnamefont {Hage}}, \bibinfo {author} {\bibfnamefont
  {H.-A.}\ \bibnamefont {Bachor}},\ and\ \bibinfo {author} {\bibfnamefont
  {W.~P.}\ \bibnamefont {Bowen}},\ }\bibfield  {title} {\bibinfo {title}
  {Biological measurement beyond the quantum limit},\ }\href
  {https://www.nature.com/articles/nphoton.2012.346} {\bibfield  {journal}
  {\bibinfo  {journal} {Nature Photonics}\ }\textbf {\bibinfo {volume} {7}},\
  \bibinfo {pages} {229} (\bibinfo {year} {2013})}\BibitemShut {NoStop}%
\bibitem [{\citenamefont {Parniak}\ \emph {et~al.}(2018)\citenamefont
  {Parniak}, \citenamefont {Bor\'owka}, \citenamefont {Boroszko}, \citenamefont
  {Wasilewski}, \citenamefont {Banaszek},\ and\ \citenamefont
  {Demkowicz-Dobrza\ifmmode~\acute{n}\else
  \'{n}\fi{}ski}}]{Parniak2018beating}%
  \BibitemOpen
  \bibfield  {author} {\bibinfo {author} {\bibfnamefont {M.}~\bibnamefont
  {Parniak}}, \bibinfo {author} {\bibfnamefont {S.}~\bibnamefont {Bor\'owka}},
  \bibinfo {author} {\bibfnamefont {K.}~\bibnamefont {Boroszko}}, \bibinfo
  {author} {\bibfnamefont {W.}~\bibnamefont {Wasilewski}}, \bibinfo {author}
  {\bibfnamefont {K.}~\bibnamefont {Banaszek}},\ and\ \bibinfo {author}
  {\bibfnamefont {R.}~\bibnamefont {Demkowicz-Dobrza\ifmmode~\acute{n}\else
  \'{n}\fi{}ski}},\ }\bibfield  {title} {\bibinfo {title} {Beating the rayleigh
  limit using two-photon interference},\ }\href
  {https://doi.org/10.1103/PhysRevLett.121.250503} {\bibfield  {journal}
  {\bibinfo  {journal} {Phys. Rev. Lett.}\ }\textbf {\bibinfo {volume} {121}},\
  \bibinfo {pages} {250503} (\bibinfo {year} {2018})}\BibitemShut {NoStop}%
\bibitem [{\citenamefont {Walborn}\ \emph {et~al.}(2003)\citenamefont
  {Walborn}, \citenamefont {de~Oliveira}, \citenamefont {P\'adua},\ and\
  \citenamefont {Monken}}]{walborn2003multimode}%
  \BibitemOpen
  \bibfield  {author} {\bibinfo {author} {\bibfnamefont {S.~P.}\ \bibnamefont
  {Walborn}}, \bibinfo {author} {\bibfnamefont {A.~N.}\ \bibnamefont
  {de~Oliveira}}, \bibinfo {author} {\bibfnamefont {S.}~\bibnamefont
  {P\'adua}},\ and\ \bibinfo {author} {\bibfnamefont {C.~H.}\ \bibnamefont
  {Monken}},\ }\bibfield  {title} {\bibinfo {title} {Multimode hong-ou-mandel
  interference},\ }\href {https://doi.org/10.1103/PhysRevLett.90.143601}
  {\bibfield  {journal} {\bibinfo  {journal} {Phys. Rev. Lett.}\ }\textbf
  {\bibinfo {volume} {90}},\ \bibinfo {pages} {143601} (\bibinfo {year}
  {2003})}\BibitemShut {NoStop}%
\bibitem [{\citenamefont {Deng}\ \emph {et~al.}(2006)\citenamefont {Deng},
  \citenamefont {Dang},\ and\ \citenamefont {Wang}}]{deng2006spatial}%
  \BibitemOpen
  \bibfield  {author} {\bibinfo {author} {\bibfnamefont {L.-P.}\ \bibnamefont
  {Deng}}, \bibinfo {author} {\bibfnamefont {G.-F.}\ \bibnamefont {Dang}},\
  and\ \bibinfo {author} {\bibfnamefont {K.}~\bibnamefont {Wang}},\ }\bibfield
  {title} {\bibinfo {title} {Spatial-mode two-photon interference at a beam
  splitter},\ }\href {https://doi.org/10.1103/PhysRevA.74.063819} {\bibfield
  {journal} {\bibinfo  {journal} {Phys. Rev. A}\ }\textbf {\bibinfo {volume}
  {74}},\ \bibinfo {pages} {063819} (\bibinfo {year} {2006})}\BibitemShut
  {NoStop}%
\bibitem [{\citenamefont {T{\"o}ppel}\ \emph {et~al.}(2012)\citenamefont
  {T{\"o}ppel}, \citenamefont {Aiello},\ and\ \citenamefont
  {Leuchs}}]{toppel2012all}%
  \BibitemOpen
  \bibfield  {author} {\bibinfo {author} {\bibfnamefont {F.}~\bibnamefont
  {T{\"o}ppel}}, \bibinfo {author} {\bibfnamefont {A.}~\bibnamefont {Aiello}},\
  and\ \bibinfo {author} {\bibfnamefont {G.}~\bibnamefont {Leuchs}},\
  }\bibfield  {title} {\bibinfo {title} {All photons are equal but some photons
  are more equal than others},\ }\href {https://10.1088/1367-2630/14/9/093051}
  {\bibfield  {journal} {\bibinfo  {journal} {New Journal of Physics}\ }\textbf
  {\bibinfo {volume} {14}},\ \bibinfo {pages} {093051} (\bibinfo {year}
  {2012})}\BibitemShut {NoStop}%
\bibitem [{\citenamefont {Cui}\ \emph {et~al.}(2023{\natexlab{a}})\citenamefont
  {Cui}, \citenamefont {Yi},\ and\ \citenamefont {Yang}}]{cui2023quantum}%
  \BibitemOpen
  \bibfield  {author} {\bibinfo {author} {\bibfnamefont {D.}~\bibnamefont
  {Cui}}, \bibinfo {author} {\bibfnamefont {X.}~\bibnamefont {Yi}},\ and\
  \bibinfo {author} {\bibfnamefont {L.-P.}\ \bibnamefont {Yang}},\ }\bibfield
  {title} {\bibinfo {title} {Quantum imaging exploiting twisted photon pairs},\
  }\href {https://doi.org/https://doi.org/10.1002/qute.202370053} {\bibfield
  {journal} {\bibinfo  {journal} {Advanced Quantum Technologies}\ }\textbf
  {\bibinfo {volume} {6}},\ \bibinfo {pages} {2370053} (\bibinfo {year}
  {2023}{\natexlab{a}})}\BibitemShut {NoStop}%
\bibitem [{\citenamefont {Yang}\ and\ \citenamefont
  {Jacob}(2021)}]{yang2021quantum}%
  \BibitemOpen
  \bibfield  {author} {\bibinfo {author} {\bibfnamefont {L.-P.}\ \bibnamefont
  {Yang}}\ and\ \bibinfo {author} {\bibfnamefont {Z.}~\bibnamefont {Jacob}},\
  }\bibfield  {title} {\bibinfo {title} {Non-classical photonic spin texture of
  quantum structured light},\ }\href
  {https://doi.org/10.1038/s42005-021-00726-w} {\bibfield  {journal} {\bibinfo
  {journal} {Communications Physics}\ }\textbf {\bibinfo {volume} {4}},\
  \bibinfo {pages} {221} (\bibinfo {year} {2021})}\BibitemShut {NoStop}%
\bibitem [{\citenamefont {Hong}\ and\ \citenamefont
  {Mandel}(1985)}]{Hong1985SPDC}%
  \BibitemOpen
  \bibfield  {author} {\bibinfo {author} {\bibfnamefont {C.~K.}\ \bibnamefont
  {Hong}}\ and\ \bibinfo {author} {\bibfnamefont {L.}~\bibnamefont {Mandel}},\
  }\bibfield  {title} {\bibinfo {title} {Theory of parametric frequency down
  conversion of light},\ }\href {https://doi.org/10.1103/PhysRevA.31.2409}
  {\bibfield  {journal} {\bibinfo  {journal} {Phys. Rev. A}\ }\textbf {\bibinfo
  {volume} {31}},\ \bibinfo {pages} {2409} (\bibinfo {year}
  {1985})}\BibitemShut {NoStop}%
\bibitem [{\citenamefont {Monken}\ \emph {et~al.}(1998)\citenamefont {Monken},
  \citenamefont {Ribeiro},\ and\ \citenamefont {P\'adua}}]{Monken1998SPDC}%
  \BibitemOpen
  \bibfield  {author} {\bibinfo {author} {\bibfnamefont {C.~H.}\ \bibnamefont
  {Monken}}, \bibinfo {author} {\bibfnamefont {P.~H.~S.}\ \bibnamefont
  {Ribeiro}},\ and\ \bibinfo {author} {\bibfnamefont {S.}~\bibnamefont
  {P\'adua}},\ }\bibfield  {title} {\bibinfo {title} {Transfer of angular
  spectrum and image formation in spontaneous parametric down-conversion},\
  }\href {https://doi.org/10.1103/PhysRevA.57.3123} {\bibfield  {journal}
  {\bibinfo  {journal} {Phys. Rev. A}\ }\textbf {\bibinfo {volume} {57}},\
  \bibinfo {pages} {3123} (\bibinfo {year} {1998})}\BibitemShut {NoStop}%
\bibitem [{\citenamefont {Black}\ \emph {et~al.}(2019)\citenamefont {Black},
  \citenamefont {Giese}, \citenamefont {Braverman}, \citenamefont {Zollo},
  \citenamefont {Barnett},\ and\ \citenamefont {Boyd}}]{Black2019nonlocal}%
  \BibitemOpen
  \bibfield  {author} {\bibinfo {author} {\bibfnamefont {A.~N.}\ \bibnamefont
  {Black}}, \bibinfo {author} {\bibfnamefont {E.}~\bibnamefont {Giese}},
  \bibinfo {author} {\bibfnamefont {B.}~\bibnamefont {Braverman}}, \bibinfo
  {author} {\bibfnamefont {N.}~\bibnamefont {Zollo}}, \bibinfo {author}
  {\bibfnamefont {S.~M.}\ \bibnamefont {Barnett}},\ and\ \bibinfo {author}
  {\bibfnamefont {R.~W.}\ \bibnamefont {Boyd}},\ }\bibfield  {title} {\bibinfo
  {title} {Quantum nonlocal aberration cancellation},\ }\href
  {https://doi.org/10.1103/PhysRevLett.123.143603} {\bibfield  {journal}
  {\bibinfo  {journal} {Phys. Rev. Lett.}\ }\textbf {\bibinfo {volume} {123}},\
  \bibinfo {pages} {143603} (\bibinfo {year} {2019})}\BibitemShut {NoStop}%
\bibitem [{\citenamefont {Pittman}\ \emph {et~al.}(1995)\citenamefont
  {Pittman}, \citenamefont {Shih}, \citenamefont {Strekalov},\ and\
  \citenamefont {Sergienko}}]{Pittman1995}%
  \BibitemOpen
  \bibfield  {author} {\bibinfo {author} {\bibfnamefont {T.~B.}\ \bibnamefont
  {Pittman}}, \bibinfo {author} {\bibfnamefont {Y.~H.}\ \bibnamefont {Shih}},
  \bibinfo {author} {\bibfnamefont {D.~V.}\ \bibnamefont {Strekalov}},\ and\
  \bibinfo {author} {\bibfnamefont {A.~V.}\ \bibnamefont {Sergienko}},\
  }\bibfield  {title} {\bibinfo {title} {Optical imaging by means of two-photon
  quantum entanglement},\ }\href {https://doi.org/10.1103/PhysRevA.52.R3429}
  {\bibfield  {journal} {\bibinfo  {journal} {Phys. Rev. A}\ }\textbf {\bibinfo
  {volume} {52}},\ \bibinfo {pages} {R3429} (\bibinfo {year}
  {1995})}\BibitemShut {NoStop}%
\bibitem [{\citenamefont {Cui}\ \emph {et~al.}(2023{\natexlab{b}})\citenamefont
  {Cui}, \citenamefont {Wang}, \citenamefont {Yi},\ and\ \citenamefont
  {Yang}}]{supplementary}%
  \BibitemOpen
  \bibfield  {author} {\bibinfo {author} {\bibfnamefont {D.}~\bibnamefont
  {Cui}}, \bibinfo {author} {\bibfnamefont {X.-L.}\ \bibnamefont {Wang}},
  \bibinfo {author} {\bibfnamefont {X.~X.}\ \bibnamefont {Yi}},\ and\ \bibinfo
  {author} {\bibfnamefont {L.-P.}\ \bibnamefont {Yang}},\ }\href@noop {} {\emph
  {\bibinfo {title} {Supplementary for "Control of quantum coherence of photons
  exploiting quantum entanglement"}}} (\bibinfo {year}
  {2023}{\natexlab{b}})\BibitemShut {NoStop}%
\bibitem [{\citenamefont {Zhang}\ \emph {et~al.}(2016)\citenamefont {Zhang},
  \citenamefont {Roux}, \citenamefont {Konrad}, \citenamefont {Agnew},
  \citenamefont {Leach},\ and\ \citenamefont {Forbes}}]{zhang2016engineering}%
  \BibitemOpen
  \bibfield  {author} {\bibinfo {author} {\bibfnamefont {Y.}~\bibnamefont
  {Zhang}}, \bibinfo {author} {\bibfnamefont {F.~S.}\ \bibnamefont {Roux}},
  \bibinfo {author} {\bibfnamefont {T.}~\bibnamefont {Konrad}}, \bibinfo
  {author} {\bibfnamefont {M.}~\bibnamefont {Agnew}}, \bibinfo {author}
  {\bibfnamefont {J.}~\bibnamefont {Leach}},\ and\ \bibinfo {author}
  {\bibfnamefont {A.}~\bibnamefont {Forbes}},\ }\bibfield  {title} {\bibinfo
  {title} {Engineering two-photon high-dimensional states through quantum
  interference},\ }\href
  {https://www.science.org/doi/full/10.1126/sciadv.1501165} {\bibfield
  {journal} {\bibinfo  {journal} {Science advances}\ }\textbf {\bibinfo
  {volume} {2}},\ \bibinfo {pages} {e1501165} (\bibinfo {year}
  {2016})}\BibitemShut {NoStop}%
\bibitem [{\citenamefont {D'Ambrosio}\ \emph {et~al.}(2019)\citenamefont
  {D'Ambrosio}, \citenamefont {Carvacho}, \citenamefont {Agresti},
  \citenamefont {Marrucci},\ and\ \citenamefont
  {Sciarrino}}]{Ambrosio2019tunable}%
  \BibitemOpen
  \bibfield  {author} {\bibinfo {author} {\bibfnamefont {V.}~\bibnamefont
  {D'Ambrosio}}, \bibinfo {author} {\bibfnamefont {G.}~\bibnamefont
  {Carvacho}}, \bibinfo {author} {\bibfnamefont {I.}~\bibnamefont {Agresti}},
  \bibinfo {author} {\bibfnamefont {L.}~\bibnamefont {Marrucci}},\ and\
  \bibinfo {author} {\bibfnamefont {F.}~\bibnamefont {Sciarrino}},\ }\bibfield
  {title} {\bibinfo {title} {Tunable two-photon quantum interference of
  structured light},\ }\href {https://doi.org/10.1103/PhysRevLett.122.013601}
  {\bibfield  {journal} {\bibinfo  {journal} {Phys. Rev. Lett.}\ }\textbf
  {\bibinfo {volume} {122}},\ \bibinfo {pages} {013601} (\bibinfo {year}
  {2019})}\BibitemShut {NoStop}%
\bibitem [{\citenamefont {Liu}\ \emph {et~al.}(2022)\citenamefont {Liu},
  \citenamefont {Chen}, \citenamefont {Xu}, \citenamefont {Cheng},
  \citenamefont {Ren}, \citenamefont {Dong}, \citenamefont {Lou}, \citenamefont
  {Yang}, \citenamefont {Xue}, \citenamefont {Liu}, \citenamefont {Zhu},
  \citenamefont {Wang},\ and\ \citenamefont {Wang}}]{liu2022hong}%
  \BibitemOpen
  \bibfield  {author} {\bibinfo {author} {\bibfnamefont {Z.-F.}\ \bibnamefont
  {Liu}}, \bibinfo {author} {\bibfnamefont {C.}~\bibnamefont {Chen}}, \bibinfo
  {author} {\bibfnamefont {J.-M.}\ \bibnamefont {Xu}}, \bibinfo {author}
  {\bibfnamefont {Z.-M.}\ \bibnamefont {Cheng}}, \bibinfo {author}
  {\bibfnamefont {Z.-C.}\ \bibnamefont {Ren}}, \bibinfo {author} {\bibfnamefont
  {B.-W.}\ \bibnamefont {Dong}}, \bibinfo {author} {\bibfnamefont {Y.-C.}\
  \bibnamefont {Lou}}, \bibinfo {author} {\bibfnamefont {Y.-X.}\ \bibnamefont
  {Yang}}, \bibinfo {author} {\bibfnamefont {S.-T.}\ \bibnamefont {Xue}},
  \bibinfo {author} {\bibfnamefont {Z.-H.}\ \bibnamefont {Liu}}, \bibinfo
  {author} {\bibfnamefont {W.-Z.}\ \bibnamefont {Zhu}}, \bibinfo {author}
  {\bibfnamefont {X.-L.}\ \bibnamefont {Wang}},\ and\ \bibinfo {author}
  {\bibfnamefont {H.-T.}\ \bibnamefont {Wang}},\ }\bibfield  {title} {\bibinfo
  {title} {Hong-ou-mandel interference between two hyperentangled photons
  enables observation of symmetric and antisymmetric particle exchange
  phases},\ }\href {https://doi.org/10.1103/PhysRevLett.129.263602} {\bibfield
  {journal} {\bibinfo  {journal} {Phys. Rev. Lett.}\ }\textbf {\bibinfo
  {volume} {129}},\ \bibinfo {pages} {263602} (\bibinfo {year}
  {2022})}\BibitemShut {NoStop}%
\bibitem [{\citenamefont {Maga\~na Loaiza}\ \emph {et~al.}(2014)\citenamefont
  {Maga\~na Loaiza}, \citenamefont {Mirhosseini}, \citenamefont {Rodenburg},\
  and\ \citenamefont {Boyd}}]{maga2014amplification}%
  \BibitemOpen
  \bibfield  {author} {\bibinfo {author} {\bibfnamefont {O.~S.}\ \bibnamefont
  {Maga\~na Loaiza}}, \bibinfo {author} {\bibfnamefont {M.}~\bibnamefont
  {Mirhosseini}}, \bibinfo {author} {\bibfnamefont {B.}~\bibnamefont
  {Rodenburg}},\ and\ \bibinfo {author} {\bibfnamefont {R.~W.}\ \bibnamefont
  {Boyd}},\ }\bibfield  {title} {\bibinfo {title} {Amplification of angular
  rotations using weak measurements},\ }\href
  {https://doi.org/10.1103/PhysRevLett.112.200401} {\bibfield  {journal}
  {\bibinfo  {journal} {Phys. Rev. Lett.}\ }\textbf {\bibinfo {volume} {112}},\
  \bibinfo {pages} {200401} (\bibinfo {year} {2014})}\BibitemShut {NoStop}%
\bibitem [{\citenamefont {Yang}(2023)}]{Yang2023Geometric}%
  \BibitemOpen
  \bibfield  {author} {\bibinfo {author} {\bibfnamefont {L.-P.}\ \bibnamefont
  {Yang}},\ }\bibfield  {title} {\bibinfo {title} {Geometric phase for twisted
  light},\ }\href {https://doi.org/10.1364/OE.476989} {\bibfield  {journal}
  {\bibinfo  {journal} {Opt. Express}\ }\textbf {\bibinfo {volume} {31}},\
  \bibinfo {pages} {10287} (\bibinfo {year} {2023})}\BibitemShut {NoStop}%
\bibitem [{\citenamefont {Alexeyev}\ and\ \citenamefont
  {Yavorsky}(2006{\natexlab{a}})}]{alexeyev2006topological}%
  \BibitemOpen
  \bibfield  {author} {\bibinfo {author} {\bibfnamefont {C.}~\bibnamefont
  {Alexeyev}}\ and\ \bibinfo {author} {\bibfnamefont {M.}~\bibnamefont
  {Yavorsky}},\ }\bibfield  {title} {\bibinfo {title} {Topological phase
  evolving from the orbital angular momentum of ‘coiled’ quantum
  vortices},\ }\href
  {https://iopscience.iop.org/article/10.1088/1464-4258/8/9/008} {\bibfield
  {journal} {\bibinfo  {journal} {Journal of Optics A: Pure and Applied
  Optics}\ }\textbf {\bibinfo {volume} {8}},\ \bibinfo {pages} {752} (\bibinfo
  {year} {2006}{\natexlab{a}})}\BibitemShut {NoStop}%
\bibitem [{\citenamefont {Alexeyev}\ and\ \citenamefont
  {Yavorsky}(2006{\natexlab{b}})}]{Alexeyev2006}%
  \BibitemOpen
  \bibfield  {author} {\bibinfo {author} {\bibfnamefont {C.~N.}\ \bibnamefont
  {Alexeyev}}\ and\ \bibinfo {author} {\bibfnamefont {M.~A.}\ \bibnamefont
  {Yavorsky}},\ }\bibfield  {title} {\bibinfo {title} {Berry’s phase for
  optical vortices in coiled optical fibres},\ }\href
  {https://dx.doi.org/10.1088/1464-4258/9/1/002} {\bibfield  {journal}
  {\bibinfo  {journal} {Journal of Optics A: Pure and Applied Optics}\ }\textbf
  {\bibinfo {volume} {9}},\ \bibinfo {pages} {6} (\bibinfo {year}
  {2006}{\natexlab{b}})}\BibitemShut {NoStop}%
\bibitem [{\citenamefont {Chow}\ \emph {et~al.}(1985)\citenamefont {Chow},
  \citenamefont {Gea-Banacloche}, \citenamefont {Pedrotti}, \citenamefont
  {Sanders}, \citenamefont {Schleich},\ and\ \citenamefont
  {Scully}}]{Chow1985Sagnac}%
  \BibitemOpen
  \bibfield  {author} {\bibinfo {author} {\bibfnamefont {W.~W.}\ \bibnamefont
  {Chow}}, \bibinfo {author} {\bibfnamefont {J.}~\bibnamefont
  {Gea-Banacloche}}, \bibinfo {author} {\bibfnamefont {L.~M.}\ \bibnamefont
  {Pedrotti}}, \bibinfo {author} {\bibfnamefont {V.~E.}\ \bibnamefont
  {Sanders}}, \bibinfo {author} {\bibfnamefont {W.}~\bibnamefont {Schleich}},\
  and\ \bibinfo {author} {\bibfnamefont {M.~O.}\ \bibnamefont {Scully}},\
  }\bibfield  {title} {\bibinfo {title} {The ring laser gyro},\ }\href
  {https://doi.org/10.1103/RevModPhys.57.61} {\bibfield  {journal} {\bibinfo
  {journal} {Rev. Mod. Phys.}\ }\textbf {\bibinfo {volume} {57}},\ \bibinfo
  {pages} {61} (\bibinfo {year} {1985})}\BibitemShut {NoStop}%
\bibitem [{\citenamefont {Heintzmann}\ and\ \citenamefont
  {Cremer}(1999)}]{heintzmann1999laterally}%
  \BibitemOpen
  \bibfield  {author} {\bibinfo {author} {\bibfnamefont {R.}~\bibnamefont
  {Heintzmann}}\ and\ \bibinfo {author} {\bibfnamefont {C.~G.}\ \bibnamefont
  {Cremer}},\ }\bibfield  {title} {\bibinfo {title} {Laterally modulated
  excitation microscopy: improvement of resolution by using a diffraction
  grating},\ }in\ \href {https://doi.org/10.1117/12.336833} {\emph {\bibinfo
  {booktitle} {Optical biopsies and microscopic techniques III}}},\ Vol.\
  \bibinfo {volume} {3568}\ (\bibinfo {organization} {SPIE},\ \bibinfo {year}
  {1999})\ pp.\ \bibinfo {pages} {185--196}\BibitemShut {NoStop}%
\bibitem [{\citenamefont {Gustafsson}(2000)}]{gustafsson2000surpassing}%
  \BibitemOpen
  \bibfield  {author} {\bibinfo {author} {\bibfnamefont {M.~G.}\ \bibnamefont
  {Gustafsson}},\ }\bibfield  {title} {\bibinfo {title} {Surpassing the lateral
  resolution limit by a factor of two using structured illumination
  microscopy},\ }\href {https://doi.org/10.1046/j.1365-2818.2000.00710.x}
  {\bibfield  {journal} {\bibinfo  {journal} {Journal of microscopy}\ }\textbf
  {\bibinfo {volume} {198}},\ \bibinfo {pages} {82} (\bibinfo {year}
  {2000})}\BibitemShut {NoStop}%
\bibitem [{\citenamefont {Loudon}(2000)}]{loudon2000quantum}%
  \BibitemOpen
  \bibfield  {author} {\bibinfo {author} {\bibfnamefont {R.}~\bibnamefont
  {Loudon}},\ }\href@noop {} {\emph {\bibinfo {title} {The quantum theory of
  light}}}\ (\bibinfo  {publisher} {OUP Oxford},\ \bibinfo {year} {2000})\
  \bibinfo {note} {, Chap. 6}\BibitemShut {NoStop}%
\bibitem [{\citenamefont {Ritboon}\ \emph {et~al.}(2019)\citenamefont
  {Ritboon}, \citenamefont {Croke},\ and\ \citenamefont
  {Barnett}}]{Ritboon2019Optical}%
  \BibitemOpen
  \bibfield  {author} {\bibinfo {author} {\bibfnamefont {A.}~\bibnamefont
  {Ritboon}}, \bibinfo {author} {\bibfnamefont {S.}~\bibnamefont {Croke}},\
  and\ \bibinfo {author} {\bibfnamefont {S.~M.}\ \bibnamefont {Barnett}},\
  }\bibfield  {title} {\bibinfo {title} {Optical angular momentum transfer on
  total internal reflection},\ }\href {https://doi.org/10.1364/JOSAB.36.000482}
  {\bibfield  {journal} {\bibinfo  {journal} {J. Opt. Soc. Am. B}\ }\textbf
  {\bibinfo {volume} {36}},\ \bibinfo {pages} {482} (\bibinfo {year}
  {2019})}\BibitemShut {NoStop}%
\bibitem [{\citenamefont {Bra{\'n}czyk}(2017)}]{branczyk2017hong}%
  \BibitemOpen
  \bibfield  {author} {\bibinfo {author} {\bibfnamefont {A.~M.}\ \bibnamefont
  {Bra{\'n}czyk}},\ }\bibfield  {title} {\bibinfo {title} {Hong-ou-mandel
  interference},\ }\href {https://arxiv.org/pdf/1711.00080.pdf} {\bibfield
  {journal} {\bibinfo  {journal} {arXiv:1711.00080}\ } (\bibinfo {year}
  {2017})}\BibitemShut {NoStop}%
\end{thebibliography}%


\widetext
\newpage
\begin{center}
\textbf{\large Supplementary Materials for ``Control of quantum coherence of photons exploiting quantum entanglement''}
\end{center}

\maketitle

\tableofcontents

\renewcommand{\theequation}{S\arabic{equation}}
\setcounter{figure}{0}
\renewcommand{\thefigure}{S\arabic{figure}}

In this supplementary material, we present a comprehensive theory of Hong-Ou-Mandel interference involving three-dimensional structured photon pairs and provide several typical examples.

\maketitle
\begin{appendix}

\section{Quantum theory of HOM interference for three-dimensional structured photon pairs}\label{sec2}

\begin{figure}
\includegraphics[width=9cm]{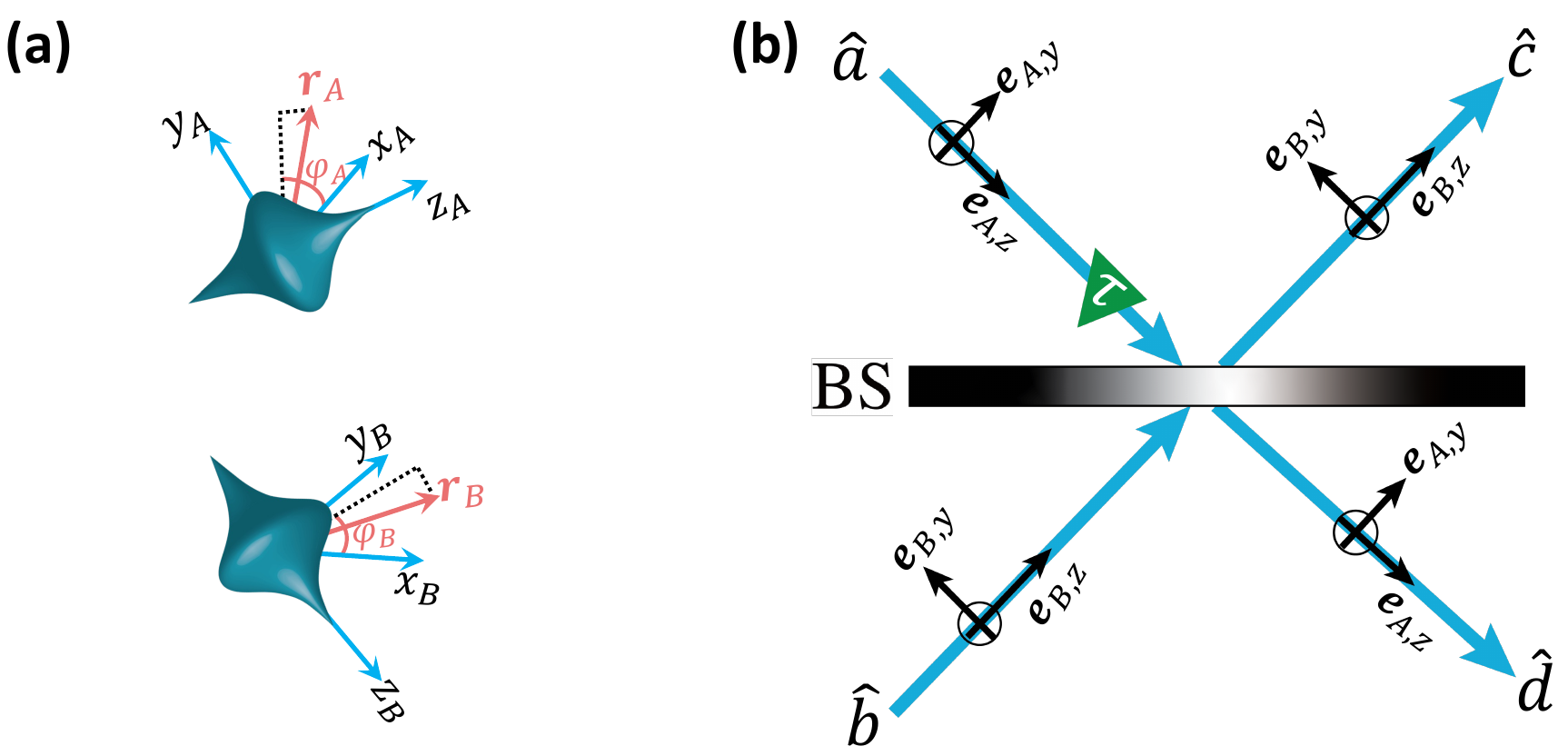}
\caption{\label{fig:1}
(a) Two-coordinate-frame formalism.
These two coordinates correspond to two propagation paths $A$ and $B$ of the two paraxial photons. The wave-packet functions of the two photons are expressed with coordinates $\boldsymbol{r}_A = x_A \boldsymbol{e}_{A,x}+y_A \boldsymbol{e}_{A,y}+z_A \boldsymbol{e}_{A,z}$ and $\boldsymbol{r}_B=x_B \boldsymbol{e}_{B,x}+y_B \boldsymbol{e}_{B,y}+z_B \boldsymbol{e}_{B,z}$, respectively. Two cylindrical coordinate frames have been employed with $x_{A(B)} = \rho_{A(B)}\cos\varphi_{A(B)}$, $y_{A(B)} = \rho_{A(B)}\sin\varphi_{A(B)}$, and $\rho_{A(B)}=\sqrt{x^2_{A(B)}+y^2_{A(B)}}$. (b) Input-output relations between ladder operators at a beam splitter (BS). The  bosonic modes
at the input ports are represented by the annihilation operators $\hat{a}$ and $\hat{b}$. The  bosonic modes at the output ports are represented by annihilation operators $\hat{c}$ and
$\hat{d}$. A time delay $\tau$ is introduced to generate the Hong-Ou-Mandel dip or peak.}
\end{figure}

This section presents the general theory of HOM interference, which involves the quantum state of a two-photon pulse with polarization degrees of freedom. The state can be expanded using plane-wave modes~\cite{loudon2000quantum,yang2022quantum}
\begin{equation}
\left|P_{\xi}\right\rangle =\frac{1}{\sqrt{2}}\sum_{\lambda\lambda'}\int d\boldsymbol{k}\int d\boldsymbol{k}'\tilde{\xi}_{\lambda\lambda'}(\boldsymbol{k},\boldsymbol{k}')\hat{a}_{\boldsymbol{k}\lambda}^{\dagger}(t)\hat{a}_{\boldsymbol{k}'\lambda'}^{\dagger}(t)\left|0\right\rangle,\label{eq:state1}
\end{equation}
with creation operators $\hat{a}^{\dagger}_{\boldsymbol{k},\lambda}(t)=\hat{a}^{\dagger}_{\boldsymbol{k},\lambda}\exp(-i\omega_{\boldsymbol{k}}t)$ that account for the free time evolution and polarization index $\lambda$. The spectrum amplitude function (SAF) $\tilde{\xi}_{\lambda\lambda'}(\boldsymbol{k},\boldsymbol{k}')$ is normalized $\sum_{\lambda\lambda'}\int d\boldsymbol{k}\int d\boldsymbol{k}'\left|\tilde{\xi}_{\lambda\lambda'}(\boldsymbol{k},\boldsymbol{k}')\right|^2=1$ and the extra factor $1/\sqrt{2}$ results from the bosonic commutation relations $[\hat{a}_{\boldsymbol{k}\lambda},\hat{a}_{\boldsymbol{k}'\lambda'}^{\dagger}]=\delta_{\lambda\lambda'}\delta (\boldsymbol{k}-\boldsymbol{k}')$. The SAF must be symmetric under particle exchange $\{\boldsymbol{k},\lambda\}\leftrightarrow \{\boldsymbol{k}',\lambda'\}$ due to the commutation relations $[\hat{a}_{\boldsymbol{k}\lambda},\hat{a}_{\boldsymbol{k}'\lambda'}]=0$ and $[\hat{a}_{\boldsymbol{k}\lambda}^{\dagger},\hat{a}_{\boldsymbol{k}'\lambda'}^{\dagger}]=0$. By introducing the effective field operator of photons, the two-photon state can be re-expressed as~\cite{yang2022quantum}
\begin{equation}
\left|P_{\xi}\right\rangle =\frac{1}{\sqrt{2}}\sum_{\lambda\lambda'}\int d\boldsymbol{r}\int d\boldsymbol{r}'\xi_{\lambda\lambda'}(\boldsymbol{r},\boldsymbol{r}',t)\hat{\psi}_{\lambda}^{\dagger}(\boldsymbol{r})\hat{\psi}_{\lambda'}^{\dagger}(\boldsymbol{r}')\left|0\right\rangle,\label{eq:state2}
\end{equation}
with commutation relation $[\hat{\psi}_{\lambda}(\boldsymbol{r}),\hat{\psi}_{\lambda'}^{\dagger}(\boldsymbol{r}')]=\delta_{\lambda,\lambda'}\delta(\boldsymbol{r}-\boldsymbol{r}')$. The real-space WPF
\begin{equation}
\xi_{\lambda,\lambda^{\prime}}(\boldsymbol{r},\boldsymbol{r}^{\prime},t)\!=\!\frac{1}{(2\pi)^{3}}\int\!\! d\boldsymbol{k}\!\int\!\! d\boldsymbol{k}'\tilde{\xi}_{\lambda,\lambda^{\prime}}(\boldsymbol{k},\boldsymbol{k}^{\prime})e^{i(\boldsymbol{k}\cdot\boldsymbol{r}-\omega_{\boldsymbol{k}}t+\boldsymbol{k}^{\prime}\cdot\boldsymbol{r}^{\prime}-\omega_{\boldsymbol{k}'}t)}, \label{eq:xi_RS}
\end{equation}
is normalized and must also be symmetric under exchange $\{\boldsymbol{r},\lambda\}\leftrightarrow \{\boldsymbol{r}',\lambda'\}$.

Usually, paraxial photon pulses are used in experiments. Two paraxial photons propagating in different directions determined by their center wave vectors as shown in Fig.~\ref{fig:1} are spatially distinguishable since there is almost no overlap between WPFs (SAFs) of the two pulses $\xi_{\lambda\lambda'}(\boldsymbol{r},\boldsymbol{r})\approx 0$ ( $\tilde{\xi}_{\lambda\lambda'}(\boldsymbol{k},\boldsymbol{k})\approx 0$). For convenience, we can employ two coordinate frames with path labels $A$ and $B$ [See Fig.~\ref{fig:1} (a)] to expand the WPF of each pulse in its respective frame~\cite{walborn2003multimode}. The two pulses can be approximately treated as two spatially independent modes. Consequently, we can introduce two sets of independent bosonic operators to express the quantum state of two-photon pulses~\cite{cui2023quantum}
\begin{align}
\left|P_{\xi}\right\rangle & =\sum_{\lambda \lambda^{\prime}} \int d\boldsymbol{k}_A\int d\boldsymbol{k}^{\prime}_B\tilde{\xi}_{\lambda \lambda^{\prime}}(\boldsymbol{k}_A,\boldsymbol{k}^{\prime}_B)\hat{a}_{\boldsymbol{k}_{A},\lambda}^{\dagger}(t)\hat{b}_{\boldsymbol{k}_{B}^{\prime},\lambda^{\prime}}^{\dagger}(t)\left|0\right\rangle  = \sum_{\lambda \lambda^{\prime}} \!\!\int \!\!d\boldsymbol{r}_A\!\!\int\!\! d\boldsymbol{r}^{\prime}_B\xi_{\lambda \lambda^{\prime}}(\boldsymbol{r}_A,\boldsymbol{r}^{\prime}_B,t)\hat{\psi}_{a,\lambda}^{\dagger}(\boldsymbol{r}_{A})\hat{\psi}_{b,\lambda^{\prime}}^{\dagger}(\boldsymbol{r}_{B}^{\prime})\left|0\right\rangle. \label{eq:P_RS}   
\end{align} 
Here, the ladder (field) operators of two photons commute with each other, i.e., $[\hat{a}_{\boldsymbol{k}_A,\lambda},\hat{b}_{\boldsymbol{k}_B,\lambda'}^{\dagger}] = 0$ and $[\hat{\psi}_{a,\lambda}(\boldsymbol{r}_A),\hat{\psi}_{b,\lambda'}^{\dagger}(\boldsymbol{r}_B^{\prime})] = 0$ and the factor $1/\sqrt{2}$ in Eq.~(\ref{eq:xi_RS}) is removed.  In this representation, the SAF $\tilde{\xi}_{\lambda\lambda'}(\boldsymbol{k}_A,\boldsymbol{k}'_B)$ and the real-space 
WPF $\xi_{\lambda\lambda'}(\boldsymbol{r}_A,\boldsymbol{r}'_B,t)$ are not required to be exchange-symmetric any more. This enables the generation of photon pairs with both exchange symmetric and anti-symmetric WPFs~\cite{walborn2003multimode,loudon2000quantum,liu2022hong}. 

The principle of identity plays an essential role in HOM interference when two single-photon pulses meet at a beam splitter [see Fig.~\ref{fig:1} (b)]. The will be a large overlap between the WPFs of two photons in this case. The effect of indistinguishability manifests in the input-output relations for a beam splitter~\cite{walborn2003multimode,cui2023quantum},
\begin{align}
\hat{c}_{\boldsymbol{k}_{B},\lambda} & =\left(R_{\boldsymbol{k}}\hat{a}_{\bar{\boldsymbol{k}}_{A},\lambda}+T_{\boldsymbol{k}}\hat{b}_{\boldsymbol{k}_{B},\lambda}\right),\\
\hat{d}_{\boldsymbol{k}_{A},\lambda} & =\left(T_{\boldsymbol{k}}\hat{a}_{\boldsymbol{k}_{A},\lambda}+R_{\boldsymbol{k}}\hat{b}_{\bar{\boldsymbol{k}}_{B},\lambda}\right),
\end{align}
with $\bar{\boldsymbol{k}}=(k_{x},-k_{y},k_{z})$ (see the supplementary of Ref.~\cite{cui2023quantum}). During an HOM interference, two plane-wave modes $\hat{a}_{\boldsymbol{k}_A,\lambda}$ and $\hat{b}_{\boldsymbol{k}_B,\lambda}$ from different input channels could be transformed into modes $\hat{d}_{\boldsymbol{k}_A,\lambda}$ and $\hat{d}_{\bar{\boldsymbol{k}}_A,\lambda}$ in the same output channel. The commutation relation between the transmitted and the reflected photon modes at the same output port is given by $[\hat{d}_{\boldsymbol{k}_A,\lambda},\hat{d}_{\bar{\boldsymbol{k}}'_A,\lambda}^{\dagger}]=\delta (\boldsymbol{k}_A-\bar{\boldsymbol{k}}'_A)$. This implies that the principle of identity ensures that one cannot distinguish between reflected and transmitted photons in a pulse at the same output port. The operators of two different output modes commute. Photons in different output channels are still distinguishable. In the two-coordinate-frame formalism, the $y$-component of wave vector changes sign after a reflection~\cite{walborn2003multimode,cui2023quantum}. This leads to a significant effect that the sign of the quantum number of photonic OAM is inverted (i.e., $m\rightarrow -m$) in each reflection~\cite{Ritboon2019Optical}.

In the HOM interference of a paraxial photon pair, both the reflection and transmission coefficients can be approximately taken as $\boldsymbol{k}$-independent constants $T_{k}=1/\sqrt{2}$ and
$R_{k}=i/\sqrt{2}$. The output state after HOM interference is given by
\begin{align}
\left|\Psi_{{\rm out}}\right\rangle
=&\frac{1}{2} \sum_{\lambda,\lambda^{\prime}} \left[i\int d \boldsymbol{k}_A \int d \boldsymbol{k}^{\prime}_A \tilde{\xi}_{\lambda,\lambda^{\prime}}\left(\boldsymbol{k}_A, \boldsymbol{k}^{\prime}_A\right) \hat{d}_{\boldsymbol{k}_{A},\lambda}^{\dagger} (t)\hat{d}_{\overline{\boldsymbol{k}}_{A}^{\prime},\lambda^{\prime}}^{\dagger}(t)+i \int d \boldsymbol{k}_B \int d \boldsymbol{k}^{\prime}_B \tilde{\xi}_{\lambda,\lambda^{\prime}}\left(\boldsymbol{k}_B, \boldsymbol{k}^{\prime}_B\right) \hat{c}_{\overline{\boldsymbol{k}}_{B},\lambda}^{\dagger} (t)\hat{c}_{\boldsymbol{k}_{B}^{\prime},\lambda^{\prime}}^{\dagger}(t)\right . \nonumber \\ 
& \left . +\int d \boldsymbol{k}_A \int d \boldsymbol{k}^{\prime}_B \tilde{\xi}_{\lambda,\lambda^{\prime}}\left(\boldsymbol{k}_A, \boldsymbol{k}^{\prime}_B\right) \hat{d}_{\boldsymbol{k}_{A},\lambda}^{\dagger} (t)\hat{c}_{\boldsymbol{k}_{B}^{\prime},\lambda^{\prime}}^{\dagger}(t)-\int d \boldsymbol{k}_B \int d \boldsymbol{k}^{\prime}_A\tilde{\xi}_{\lambda,\lambda^{\prime}}\left(\boldsymbol{k}_B, \boldsymbol{k}^{\prime}_A\right) \hat{d}_{\overline{\boldsymbol{k}}_{A}^{\prime},\lambda^{\prime}}^{\dagger} (t)\hat{c}_{\overline{\boldsymbol{k}}_{B},\lambda}^{\dagger}(t)\right]|0\rangle, \\
=&  \frac{1}{2}\sum_{\lambda,\lambda^{\prime}}\left[i\int d\boldsymbol{r}_A\int d\boldsymbol{r}^{\prime}_A \xi_{\lambda,\lambda^{\prime}}(\boldsymbol{r}_A,\bar{\boldsymbol{r}}_A^{\prime},t)\hat{\psi}_{d,\lambda}^{\dagger}(\boldsymbol{r}_{A})\hat{\psi}_{d,\lambda^{\prime}}^{\dagger}(\boldsymbol{r}_{A}^{\prime})+i\int d\boldsymbol{r}_B\int d\boldsymbol{r}^{\prime}_B\xi_{\lambda,\lambda^{\prime}}(\bar{\boldsymbol{r}}_B,\boldsymbol{r}^{\prime}_B,t)\hat{\psi}_{c,\lambda}^{\dagger}(\boldsymbol{r}_{B})\hat{\psi}_{c,\lambda^{\prime}}^{\dagger}(\boldsymbol{r}_{B}^{\prime}) \right . \nonumber \\
& \left .+ \int d\boldsymbol{r}_A\int d\boldsymbol{r}^{\prime}_B\xi_{cd}(\boldsymbol{r}_A,\boldsymbol{r}^{\prime}_B,t)\hat{\psi}_{d,\lambda}^{\dagger}(\boldsymbol{r}_{A})\hat{\psi}_{c,\lambda^{\prime}}^{\dagger}(\boldsymbol{r}_{B}^{\prime})\right]\left|0\right\rangle,\label{eq:Psi_out2} 
\end{align}
with $\bar{\boldsymbol{r}}=(x,-y,z)$ and
WPF
\begin{equation}
\xi_{cd}(\boldsymbol{r}_A,\boldsymbol{r}^{\prime}_B,t)=\xi_{\lambda,\lambda^{\prime}}(\boldsymbol{r}_A,\boldsymbol{r}^{\prime}_B,t)-\xi_{\lambda^{\prime},\lambda}(\bar{\boldsymbol{r}}^{\prime}_B,\bar{\boldsymbol{r}}_A,t).   
\end{equation}
The first two terms in state $|\Psi_{\rm out}\rangle$ represent photons coming out from the same port of the beam splitter---bunching photons. The third term represents the two photons coming out of different ports---anti-bunching photons.

We emphasize that the HOM interference is not solely determined by the exchange symmetry of a photon pair, but the combined exchange-reflection symmetry. We introduce a symmetry index $s=\pm 1$ for the exchange-reflection symmetry condition $\xi_{\lambda,\lambda^{\prime}}(\boldsymbol{r}_A,\boldsymbol{r}^{\prime}_B,t)=s\xi_{\lambda^{\prime},\lambda}(\bar{\boldsymbol{r}}^{\prime}_B,\bar{\boldsymbol{r}}_A,t)$. Destructive (dip) and constructive (peak) HOM interferences are obtained with $s=+1$ and $s=-1$, respectively.

In experiments, the two-photon coincidence probability
\begin{equation}
P_{ cd}^{(2)} =\bra{\Psi_{\rm out}}\hat {N}_c\otimes\hat {N}_d\ket {\Psi_{\rm out}},
\label{coincidence probability11}
\end{equation}
has been used to demonstrate the HOM interference. For 
single-photon detectors with flat frequency response, the measurements at two output ports can be described by paraxial photon number operators~\cite{branczyk2017hong,yang2022quantum}
\begin{align}
&\hat {N}_c=\!\int\!\! d\boldsymbol{k}_B\hat{c}_{\boldsymbol{k}_B,\lambda}^{\dagger}\hat{c}_{\boldsymbol{k}_B,\lambda}=\int d\boldsymbol{r}_B\hat{\psi}_{c,\lambda}^{\dagger}(\boldsymbol{r}_B)\hat{\psi}_{c,\lambda}(\boldsymbol{r}_B),\\
&\hat {N}_d=\!\int\!\! d\boldsymbol{k}_A\hat{d}_{\boldsymbol{k}_A,\lambda}^{\dagger} \hat{d}_{\boldsymbol{k}_A,\lambda}=\int d\boldsymbol{r}_A\hat{\psi}_{d,\lambda}^{\dagger}(\boldsymbol{r}_A)\hat{\psi}_{d,\lambda}(\boldsymbol{r}_A).
\label{projector1}
\end{align}
From Eqs. (\ref{coincidence probability11}-\ref{projector1}), we obtain the coincidence probability
\begin{align}
P_{cd}^{(2)} &=\frac{1}{2}-\frac{1}{4}\sum_{\lambda\lambda^{\prime}}
 \int d\boldsymbol{k}_A \int d\boldsymbol{k}^{\prime}_B\left[\tilde{\xi}^*_{\lambda,\lambda^{\prime}}\left(\boldsymbol{k}_A, \boldsymbol{k}^{\prime}_B\right) \tilde{\xi}_{\lambda^{\prime},\lambda}\left(\boldsymbol{\bar{k}}^{\prime}_B, \boldsymbol{\bar{k}}_A\right)
+\tilde{\xi}^*_{\lambda^{\prime},\lambda}\left(\boldsymbol{\bar{k}}^{\prime}_B, \boldsymbol{\bar{k}}_A\right) \tilde{\xi}_{\lambda,\lambda^{\prime}}\left(\boldsymbol{k}_A, \boldsymbol{k}^{\prime}_B\right)\right],\\
&=\frac{1}{2}-\frac{1}{4}\sum_{\lambda\lambda^{\prime}}
\int d\boldsymbol{r}_A \int d\boldsymbol{r}^{\prime}_B \left[ \xi^*_{\lambda,\lambda^{\prime}}\left(\boldsymbol{r}_A, \boldsymbol{r}^{\prime}_B,t\right) \xi_{\lambda^{\prime},\lambda}\left(\boldsymbol{\bar{r}}^{\prime}_B, \boldsymbol{\bar{r}}_A,t\right)
+\xi^*_{\lambda^{\prime},\lambda}\left(\boldsymbol{\bar{r}}^{\prime}_B, \boldsymbol{\bar{r}}_A,t\right) \xi_{\lambda,\lambda^{\prime}}\left(\boldsymbol{r}_A, \boldsymbol{r}^{\prime}_B,t\right)\right].
\end{align}

A similar theory has been presented in prior literature~\cite{deng2006spatial}, which has also been extended to mixed states~\cite{toppel2012all}. In the subsequent sections, we provide a comprehensive analysis of the distinguishability of spatiotemporal and polarization properties in HOM interference using OAM or polarization entangled photon pairs. This analysis serves as inspiration for the development of precise methods to manipulate the higher-order coherence of entangled photons in the main text.

\begin{figure}
\includegraphics[width=8.5cm]{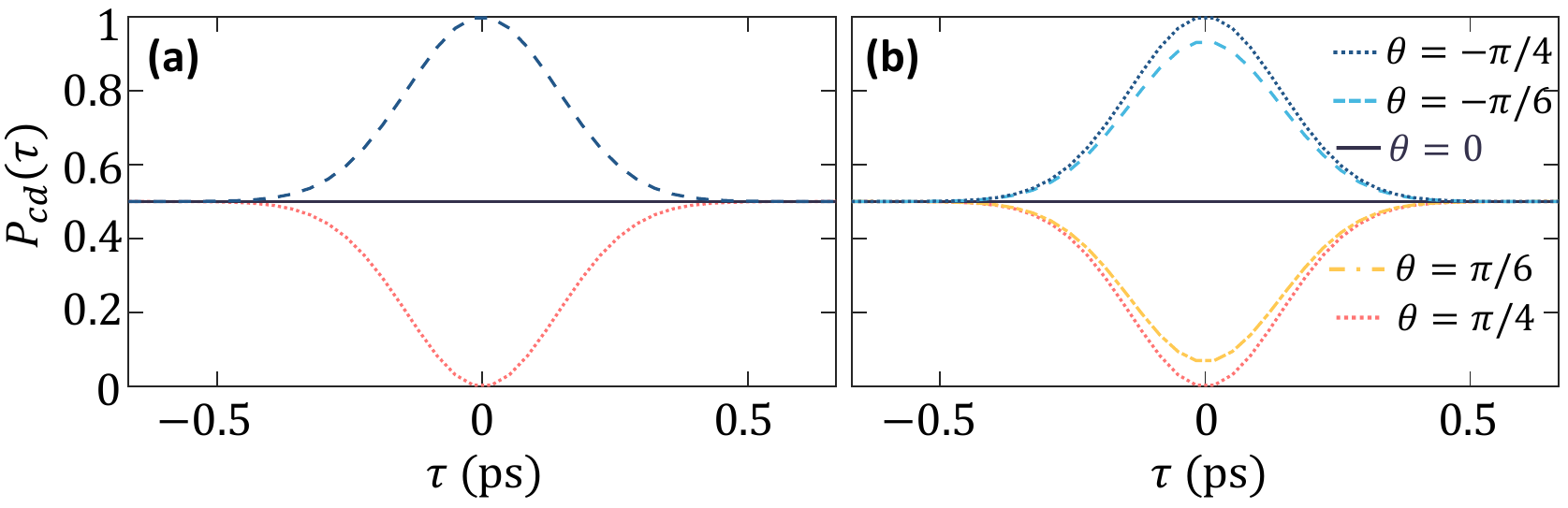}
\centering
\caption{\label{fig:2} Coincidence probability in Hong-Ou-Mandel (HOM) interference. (a) Control the HOM interference via orbital-angular-momentum degrees of photons. The pink dotted line corresponds to twisted photon pairs with spectrum amplitude functions (SAF) given in Eqs.~(\ref{eq:xi1}), (\ref{eq:entangled_1}),  and (\ref{eq:entangled_3}). The black solid line corresponds to a photon pair with SAF given in Eq.~(\ref{eq:xi2}). The navy-blue dashed line corresponds to a twisted photon pair with SAF given in Eq.~(\ref{eq:entangled_2}).
(b) Control the HOM interference via polarization degrees of photons. The curves from top to bottom correspond to $\theta = -\pi/4,-\pi/6,0,\pi/6,\pi/4$, respectively. Here, the size of the pulse is set to $3\times10^{-5}$m (corresponding to the pulse length in time $10^{-13}$ s).}
\end{figure}

\section{Spatiotemporal distinguishability}
Here we study how the spatiotemporal distinguishability of the input photons influences the HOM coincidence measurements. Specifically, we investigate the role of the helical structure of linearly polarized twisted photon pairs in HOM interference. In this subsection, we neglect the polarization indices for simplicity.

\subsection{Product-state twisted photon pair}
We first consider two input  photons in a product state with an SAF
\begin{equation}
\tilde{\xi}(\boldsymbol{k},\boldsymbol{k}^{\prime})=\mathcal{N}\tilde{\eta}(\boldsymbol{k})\tilde{\eta}(\boldsymbol{k}^{\prime})e^{i[m(\tilde{\varphi}-\tilde{\varphi}^{\prime})]}e^{-ik_{z}z_0}, \label{eq:xi1}
\end{equation}
where the  normalization factor is $\mathcal{N}=1$ and $\tilde{\varphi}$ ($\tilde{\varphi}'$) is the azimuthal angle of the wave vector $\boldsymbol{k}$ ($\boldsymbol{k}'$) in the momentum-space cylindrical coordinate. The integer $m$ represents the OAM quantum number of each photon, and the normalized function ${\tilde{\eta}(\boldsymbol{k})}$ characterizes the pulse shape and pulse length of the two photons and is usually independent on azimuthal angle $\tilde{\varphi}$~\cite{yang2022quantum,cui2023quantum}. To simplify, we will omit the path labels $A$ and $B$ unless they are necessary. We note that two photons in this pulse have opposite OAM quantum numbers. The corresponding WPF is given by
\begin{equation}
\xi(\boldsymbol{r},\boldsymbol{r}',t) =\mathcal{N}\eta_m (\boldsymbol{r},t+\tau)\eta_{-m}(\boldsymbol{r}',t) e^{im(\varphi-\varphi')}, \label{eq:eta-tilde1}
\end{equation}
with
\begin{equation}
\eta_{\pm m}(\boldsymbol{r},t)\!=\!\frac{i^{\pm m}}{\sqrt{2\pi}}\!\int_{-\infty}^{\infty}\!\!dk_{z}\!\!\int_{0}^{\infty}\!\!\tilde{\rho}d\tilde{\rho}\tilde{\eta}(k_{z},\tilde{\rho})J_{\pm m}(\rho\tilde{\rho})e^{i(k_{z}z-\omega_{\boldsymbol{k}}t)},
\end{equation}
where $\rho=\sqrt{x^2+y^2}$ and $\tilde{\rho}=\sqrt{k^2_x+k^2_y}$  represent the magnitudes of the vector projections of $\boldsymbol{r}$ and $\boldsymbol{k}$ onto the $xy$-plane respectively, and $J_n(x)$ is the $n$th order Bessel function of the first kind. We have used the fact that $J_{m}(x)=(-1)^{m}J_{-m}(x)$ and  in
the last step. We note that the function $\eta_{m}(\boldsymbol{r},t)$ is independent on the azimuthal angle $\varphi$ and $\eta_{m}(\boldsymbol{r},t)=\eta_{-m}(\boldsymbol{r},t)$~\cite{cui2023quantum}. A time delay $\tau= z_0/c$ is added in path A (see Fig.~\ref{fig:1}) (b). 

The coincidence probability after the BS is obtained from Eq.~(\ref{coincidence probability11}) as
\begin{equation}
P_{cd}^{(2)} = \frac{1}{2}\!-\!\frac{1}{2}\!\int\! d\boldsymbol{k}\!\!\int\! d\boldsymbol{k}^{\prime}\left|\tilde{\eta}(\boldsymbol{k})\right|^{2}\left|\tilde{\eta}(\boldsymbol{k}^{\prime})\right|^{2}e^{i(k_{z}-k_{z}^{\prime})z_0}.
\label{P1}
\end{equation}
In the absence of delay (i.e., $\tau=z_0/c=0$), one can easily verify that the WPF in Eq.~(\ref{eq:eta-tilde1}) satisfies the exchange-reflection condition $\xi(\bar{\boldsymbol{r}}',\bar{\boldsymbol{r}},t)=\xi(\boldsymbol{r},\boldsymbol{r}',t)$ (note the sign change in $m$ by $\bar{\boldsymbol{r}}$~\cite{cui2023quantum}). No coincidence events will be observed in experiments. An HOM interference dip will be obtained by varying the optical path (i.e., the delay $\tau$) in one of
the input ports as shown by the pink dotted line in Fig.~\ref{fig:2} (a).

We now consider two photons with the same OAM quantum number. The SAF of the photon pair 
\begin{equation}
\tilde{\xi}(\boldsymbol{k},\boldsymbol{k}^{\prime})=\mathcal{N}\tilde{\eta}(\boldsymbol{k})\tilde{\eta}(\boldsymbol{k}^{\prime})e^{i[m(\tilde{\varphi}+\tilde{\varphi}^{\prime})]}e^{-ik_{z}z_0}, \label{eq:xi2}
\end{equation}
is exchange symmetric in the absence of a delay (i.e., $z_0 =0$). However, the corresponding WPF 
\begin{equation}
\xi(\boldsymbol{r},\boldsymbol{r}',t) =\mathcal{N}\eta_m (\boldsymbol{r},t+\tau)\eta_{m}(\boldsymbol{r}',t) e^{im(\varphi+\varphi')}, \label{eq:eta-tilde2}
\end{equation}
does not have the exchange-reflection symmetry except for $m=0$~\cite{cui2023quantum}, i.e., $\xi(\bar{\boldsymbol{r}}',\bar{\boldsymbol{r}},t)\neq \pm\xi(\boldsymbol{r},\boldsymbol{r}',t)$. After the HOM interference, the coincidence probability is given by
\begin{align}
P_{cd}^{(2)} = & \frac{1}{2}-\frac{1}{2}\int d\boldsymbol{k}\int d\boldsymbol{k}^{\prime}\mathcal{N}^2|\tilde{\eta}(\boldsymbol{k})|^2|\tilde{\eta}(\boldsymbol{k}^{\prime})|^2\cos{\left[2m\left(\tilde{\varphi}+\tilde{\varphi}^{\prime}\right)\right]}e^{i(k_{z}-k_{z}^{\prime})z_0}. \label{eq:Pcd4mm}
\end{align}
We can further verify that $P^{(2)}_{cd} =\left(1-\delta_{m, 0}\right)/2$ for $z_0 = 0$. As shown by the black solid line in Fig.~\ref{fig:2} (a),
the HOM dip vanishes for  nonzero OAM quantum number
($m \neq 0$). Here, we clearly show that not the exchange symmetry but the combined exchange-reflection symmetry plays an essential role in HOM interference.

\subsection{Entangled twisted photon pair}
Next, we study the HOM interference of entangled twisted photon pairs, which have been extensively used in quantum communications and quantum sensing experiments. We first consider a twisted photon pair with an exchange symmetric SAF in the absence of a delay
\begin{equation}
\tilde{\xi}(\boldsymbol{k},\boldsymbol{k}^{\prime})=\mathcal{N}\tilde{\eta}(\boldsymbol{k})\tilde{\eta}(\boldsymbol{k}^{\prime})\left[e^{im(\tilde{\varphi}+\tilde{\varphi}^{\prime})}+ e^{-im(\tilde{\varphi}+\tilde{\varphi}^{\prime})}\right]e^{-ik_{z}z_0}, \label{eq:entangled_1}
\end{equation}
where the two photons always have the
same OAM quantum number and $\mathcal{N}=[2(1+\delta_{m,0})]^{-1/2}$. The corresponding WPF
\begin{equation}
\xi(\boldsymbol{r},\boldsymbol{r}',t) =\mathcal{N}\eta_m (\boldsymbol{r},t+\tau)\eta_{m}(\boldsymbol{r}',t) \left[e^{im(\varphi+\varphi')}+e^{-im(\varphi+\varphi')}\right], \label{eq:eta-tilde3}
\end{equation}
satisfies the exchange-reflection symmetry condition $\xi(\bar{\boldsymbol{r}}',\bar{\boldsymbol{r}},t)=\xi(\boldsymbol{r},\boldsymbol{r}',t)$ for $\tau =0$. The coincidence probability is given by
\begin{align}
P_{cd}^{(2)}=&\frac{1}{2}-\frac{1}{2}\int d\boldsymbol{k}\int d\boldsymbol{k}^{\prime}\mathcal{N}^2\left|\tilde{\eta}(\boldsymbol{k})\right|^{2}\left|\tilde{\eta}(\boldsymbol{k}^{\prime})\right|^{2} \Big(2+2 \cos{\left[2m\left(\tilde{\varphi}+\tilde{\varphi}^{\prime}\right)\right]}\Big)e^{i(k_{z}-k_{z}^{\prime})z_0}.
\end{align}
The coincidence probability vanishes $P_{cd}^{(2)}=0$ for $z_0=0$. This implies that an HOM dip will be obtained as shown by the pink dotted line in Fig.~\ref{fig:2} (a).

We now consider another entangled twisted  photon pair with exchange symmetric SAF in the absence of delay
\begin{equation}
\tilde{\xi}(\boldsymbol{k},\boldsymbol{k}^{\prime})=\mathcal{N}\tilde{\eta}(\boldsymbol{k})\tilde{\eta}(\boldsymbol{k}^{\prime})\left[e^{im(\tilde{\varphi}+\tilde{\varphi}^{\prime})}-e^{-im(\tilde{\varphi}+\tilde{\varphi}^{\prime})}\right]e^{-ik_{z}z_0}, \label{eq:entangled_2}
\end{equation}
where $\mathcal{N}=1/\sqrt{2}$ $(m\neq0)$. However, the corresponding WPF
\begin{equation}
\xi(\boldsymbol{r},\boldsymbol{r}',t) =\mathcal{N}\eta_m (\boldsymbol{r},t+\tau)\eta_{m}(\boldsymbol{r}',t) \left[e^{im(\varphi+\varphi')}-e^{-im(\varphi+\varphi')}\right], \label{eq:eta-tilde4}
\end{equation}
is exchange-reflection anti-symmetric $\xi(\bar{\boldsymbol{r}}',\bar{\boldsymbol{r}},t)=-\xi(\boldsymbol{r},\boldsymbol{r}',t)$ for $\tau=z_0/c=0$~\cite{cui2023quantum}. The corresponding coincidence probability is obtained
as 
\begin{align}
P_{cd}^{(2)}=&\frac{1}{2}+\frac{1}{2}\int d\boldsymbol{k}\int d\boldsymbol{k}^{\prime}\mathcal{N}^2\left|\tilde{\eta}(\boldsymbol{k})\right|^{2}\left|\tilde{\eta}(\boldsymbol{k}^{\prime})\right|^{2}\Big(2-2 \cos{\left[2m\left(\tilde{\varphi}+\tilde{\varphi}^{\prime}\right)\right]}\Big)e^{i(k_{z}-k_{z}^{\prime})z_0}.
\end{align}
In this case, an HOM peak will be obtained as shown by the navy-blue dashed line in Fig.~\ref{fig:2} (a).

If two photons carry an equal amount of OAM but with the opposite sign, we can have entangled photon pairs with SAF
\begin{equation}
\tilde{\xi}^{\pm}(\boldsymbol{k},\boldsymbol{k}^{\prime})=\mathcal{N}\tilde{\eta}(\boldsymbol{k})\tilde{\eta}(\boldsymbol{k}^{\prime})\left[e^{im(\tilde{\varphi}-\tilde{\varphi}^{\prime})}\pm e^{-im(\tilde{\varphi}-\tilde{\varphi}^{\prime})}\right]e^{-ik_{z}z_0}. \label{eq:entangled_3}
\end{equation}
In the absence of delay, the SAF $\tilde{\xi}^{+}(\boldsymbol{k},\boldsymbol{k}^{\prime})$ and $\tilde{\xi}^{-}(\boldsymbol{k},\boldsymbol{k}^{\prime})$ are exchange symmetric and anti-symmetric, respectively. However, the corresponding two WPFs
\begin{equation}
\xi^{\pm}(\boldsymbol{r},\boldsymbol{r}',t) =\mathcal{N}\eta_m (\boldsymbol{r},t+\tau)\eta_{m}(\boldsymbol{r}',t) \left[e^{im(\varphi-\varphi')}\pm e^{-im(\varphi-\varphi')}\right], \label{eq:eta-tilde5} 
\end{equation}
both satisfy the exchange-reflection symmetric condition $\xi^{\pm} (\bar{\boldsymbol{r}}',\bar{\boldsymbol{r}},t)=\xi^{\pm}(\boldsymbol{r},\boldsymbol{r}',t)$ when $\tau=0$. The coincidence probability is given by
\begin{equation}
\begin{aligned}
P_{cd,\pm}^{(2)}=&\frac{1}{2}-\frac{1}{2}\int d\boldsymbol{k}\int d\boldsymbol{k}^{\prime}\mathcal{N}^2\left|\tilde{\eta}(\boldsymbol{k})\right|^{2}\left|\tilde{\eta}(\boldsymbol{k}^{\prime})\right|^{2}\Big(2\pm2 \cos{\left[2m\left(\tilde{\varphi}-\tilde{\varphi}^{\prime}\right)\right]}\Big)e^{i(k_{z}-k_{z}^{\prime})z_0}.
\end{aligned}
\end{equation}
Only an HOM dip will be obtained for both the exchange symmetric and antisymmetric photon pairs as shown by the pink dotted line in Fig.~\ref{fig:2} (a).

\section{Polarization distinguishability}
In this subsection, we consider the influences of polarization distinguishability on two-photon HOM coincidence measurements. We apply our theory to a photon pair with entangled polarization states
\begin{equation}
\tilde{\xi}_{\lambda\lambda'}(\boldsymbol{k},\boldsymbol{k}')=\frac{1}{\sqrt{2}}\tilde{\xi}(\boldsymbol{k},\boldsymbol{k}')\left[\delta_{\lambda,H}\delta_{\lambda', V}\cos\theta +\delta_{\lambda,V}\delta_{\lambda',H}\sin\theta\right],\label{eq:state3}
\end{equation}
where $H$ ($V$) denotes the horizontal (vertical) polarization and $\theta$ is the mixing angle between states $|HV\rangle$ and $|VH\rangle$. We can take the SAF $\tilde{\xi}(\boldsymbol{k},\boldsymbol{k}^{\prime})$ being of the product of two Gaussian functions, which is exchange-reflection symmetric without delay.


The coincidence probability 
\begin{equation}
P_{cd}^{(2)}=\frac{1}{2}-\frac{1}{4}\sin 2\theta \!\int\!\! d\boldsymbol{k}\!\int\!\! d\boldsymbol{k}'\!\left[\tilde{\xi}^*(\boldsymbol{k},\boldsymbol{k}')\tilde{\xi}(\bar{\boldsymbol{k}}',\bar{\boldsymbol{k}})+\rm{h.c.}\right], 
\end{equation}
can be continuously tuned by varying the mixing angle $\theta$ as shown in Fig.~\ref{fig:2} (b). For $\theta = - \pi/4$, the polarization of the photon pair is described by state $(|HV\rangle - |VH\rangle )/\sqrt{2}$, which is exchange anti-symmetric. Different from the OAM degrees of freedom, the polarization of photons does not change at the beam splitter. Therefore, the total WPF is exchange-reflection anti-symmetric and an HOM peak is obtained. Similarly, an HOM dip is obtained for $\theta=\pi/4$, since the corresponding polarization state $(|HV\rangle + |VH\rangle )/\sqrt{2}$ is exchange symmetric. For $\theta = 0$, the polarization of the photon pair is $|HV\rangle$. The two photons are distinguishable via their polarizations. Thus, no HOM interference will be observed as shown by the black solid line in Fig.\ref{fig:2} (b). The coincidence contrast decreases to zero by varying $|\theta|$ from $\pi/4$ to 0. Recently, Liu et al. experimentally studied 16 polarization-OAM hyperentangled two-photon states~\cite{liu2022hong}. In Appendix~\ref{appendixA}, we compared our theoretical results with this experiment. Our theory is consistent with their results.

\begin{table*}[ht]
\begin{tabular}{|m{65pt}<{\centering}|m{240pt}<{\centering}|m{15pt}<{\centering}|m{50pt}<{\centering}|m{50pt}<{\centering}|}
\hline 
      State in Ref. ~\cite{liu2022hong}   & State at the beam splitter & $s$ & Experiment & Theory \\
      \hline
      $\left|\phi^{+}\right\rangle_{12} \otimes\left|\mu^{+}\right\rangle_{12}$ & $\left(|H\rangle_1 |H\rangle_2+|V\rangle_1 |V\rangle_2\right) \otimes \left(|+m\rangle_1|-m\rangle_2 +|-m\rangle_1|+m\rangle_2\right)/2$ & +1 & dip & dip\\
      $\left|\phi^{+}\right\rangle_{12} \otimes\left|\mu^{-}\right\rangle_{12}$ & $\left(|H\rangle_1 |H\rangle_2+|V\rangle_1 |V\rangle_2\right) \otimes \left(|+m\rangle_1|-m\rangle_2 -|-m\rangle_1|+m\rangle_2\right)/2$ & +1 & dip  & dip \\
      $\left|\phi^{+}\right\rangle_{12} \otimes\left|\nu^{+}\right\rangle_{12}$ & $\left(|H\rangle_1 |H\rangle_2+|V\rangle_1 |V\rangle_2\right) \otimes \left(|+m\rangle_1|+m\rangle_2 +|-m\rangle_1|-m\rangle_2\right)/2$ & +1 & dip & dip\\
      $\left|\phi^{+}\right\rangle_{12} \otimes\left|\nu^{-}\right\rangle_{12}$& $\left(|H\rangle_1 |H\rangle_2+|V\rangle_1 |V\rangle_2\right) \otimes \left(|+m\rangle_1|+m\rangle_2 -|-m\rangle_1|-m\rangle_2\right)/2$ & -1 & peak & peak\\
      \hline
      $\left|\phi^{-}\right\rangle_{12} \otimes\left|\mu^{+}\right\rangle_{12}$& $\left(|H\rangle_1 |H\rangle_2-|V\rangle_1 |V\rangle_2\right) \otimes \left(|+m\rangle_1|-m\rangle_2 +|-m\rangle_1|+m\rangle_2\right)/2$ & +1 &dip &dip\\
      $\left|\phi^{-}\right\rangle_{12} \otimes\left|\mu^{-}\right\rangle_{12}$& $\left(|H\rangle_1 |H\rangle_2-|V\rangle_1 |V\rangle_2\right) \otimes \left(|+m\rangle_1|-m\rangle_2 -|-m\rangle_1|+m\rangle_2\right)/2$ & +1 &dip & dip \\
      $\left|\phi^{-}\right\rangle_{12} \otimes\left|\nu^{+}\right\rangle_{12}$& $\left(|H\rangle_1 |H\rangle_2-|V\rangle_1 |V\rangle_2\right) \otimes \left(|+m\rangle_1|+m\rangle_2 +|-m\rangle_1|-m\rangle_2\right)/2$ & +1 & dip &dip\\
      $\left|\phi^{-}\right\rangle_{12} \otimes\left|\nu^{-}\right\rangle_{12}$& $\left(|H\rangle_1 |H\rangle_2-|V\rangle_1 |V\rangle_2\right) \otimes \left(|+m\rangle_1|+m\rangle_2 -|-m\rangle_1|-m\rangle_2\right)/2$ & -1 & peak & peak\\
      \hline
      $\left|\psi^{+}\right\rangle_{12} \otimes\left|\mu^{+}\right\rangle_{12}$& $\left(|H\rangle_1 |V\rangle_2+|V\rangle_1 |H\rangle_2\right) \otimes \left(|+m\rangle_1|-m\rangle_2 +|-m\rangle_1|+m\rangle_2\right)/2$ & +1 &dip & dip\\
      $\left|\psi^{+}\right\rangle_{12} \otimes\left|\mu^{-}\right\rangle_{12}$& $\left(|H\rangle_1 |V\rangle_2+|V\rangle_1 |H\rangle_2\right) \otimes \left(|+m\rangle_1|-m\rangle_2 -|-m\rangle_1|+m\rangle_2\right)/2$ & +1 & dip & dip\\
      $\left|\psi^{+}\right\rangle_{12} \otimes\left|\nu^{+}\right\rangle_{12}$& $\left(|H\rangle_1 |V\rangle_2+|V\rangle_1 |H\rangle_2\right) \otimes \left(|+m\rangle_1|+m\rangle_2 +|-m\rangle_1|-m\rangle_2\right)/2$ & +1 & dip &dip\\
      $\left|\psi^{+}\right\rangle_{12} \otimes\left|\nu^{-}\right\rangle_{12}$& $\left(|H\rangle_1 |V\rangle_2+|V\rangle_1 |H\rangle_2\right) \otimes \left(|+m\rangle_1|+m\rangle_2 -|-m\rangle_1|-m\rangle_2\right)/2$ & +1 & peak & peak \\
      \hline
      $\left|\psi^{-}\right\rangle_{12} \otimes\left|\mu^{+}\right\rangle_{12}$& $\left(|H\rangle_1 |V\rangle_2-|V\rangle_1 |H\rangle_2\right) \otimes \left(|+m\rangle_1|-m\rangle_2 +|-m\rangle_1|+m\rangle_2\right)/2$ & -1 & peak & peak\\
      $\left|\psi^{-}\right\rangle_{12} \otimes\left|\mu^{-}\right\rangle_{12}$& $\left(|H\rangle_1 |V\rangle_2-|V\rangle_1 |H\rangle_2\right) \otimes \left(|+m\rangle_1|-m\rangle_2 -|-m\rangle_1|+m\rangle_2\right)/2$ & -1 &peak & peak \\
      $\left|\psi^{-}\right\rangle_{12} \otimes\left|\nu^{+}\right\rangle_{12}$& $\left(|H\rangle_1 |V\rangle_2-|V\rangle_1 |H\rangle_2\right) \otimes \left(|+m\rangle_1|+m\rangle_2 +|-m\rangle_1|-m\rangle_2\right)/2$ & -1& peak &peak\\
      $\left|\psi^{-}\right\rangle_{12} \otimes\left|\nu^{-}\right\rangle_{12}$& $\left(|H\rangle_1 |V\rangle_2-|V\rangle_1 |H\rangle_2\right) \otimes \left(|+m\rangle_1|+m\rangle_2 -|-m\rangle_1|-m\rangle_2\right)/2$ & +1 & dip& dip \\
      \hline
\end{tabular}
\caption{\label{tab:1}Comparision of experiment and our theoretical results for HOM interference of sixteen hyper-entangled Bell states.}
\end{table*}

\section{Comparison with the previous experiment}\label{appendixA}
In Table~\ref{tab:1}, we compare our theoretical results with the experiment~\cite{liu2022hong}. We note that the notation of the two-photon states in Ref.~\cite{liu2022hong} is different from this work. The times of reflection for the two optical paths are different. Thus, the true quantum states of the 16 two-photon pulses at the beam splitter are given by the second column of Table~\ref{tab:1}. From the output state~(\ref{eq:Psi_out2}), we see that the destructive or constructive HOM interference is determined by $\xi_{\lambda,\lambda^{\prime}}(\boldsymbol{r},\boldsymbol{r}^{\prime},t)=s \xi_{\lambda^{\prime},\lambda}(\bar{\boldsymbol{r}}^{\prime},\bar{\boldsymbol{r}},t)$ ($s=\pm 1$), not the exchange symmetry of the total wave function directly. The HOM dip and peak are obtained with $s=+1$ and peak $s=-1$, respectively. Our theory is consistent with the experiment~\cite{liu2022hong}.

\end{appendix}
\end{document}